\documentclass[reprint,pre,aps,floatfix]{revtex4-2}
\usepackage{graphicx}
\usepackage{amsfonts}
\usepackage{amssymb}
\usepackage{natbib}
\usepackage{xcolor}
\usepackage{amsmath}
\usepackage{enumerate}
\usepackage{dblfloatfix}
\usepackage[percent]{overpic}
\usepackage{verbatim}
\usepackage[percent]{overpic}
\usepackage[colorlinks=true, linkcolor=blue, citecolor=blue, urlcolor=blue]{hyperref}

\usepackage{ragged2e}
\makeatletter
\long\def\@makecaption#1#2{%
  \vskip\abovecaptionskip
  \noindent\justifying
  \hangindent=0.0em\hangafter=1 
  \leavevmode\normalfont#1\nobreak\hskip.5em\relax #2\par
  \vskip\belowcaptionskip}
\makeatother

\begin{document}

\title{Spectral statistics and localization properties of a C$_3$-symmetric billiard}

\author{Matic Orel and Marko Robnik}

\affiliation{CAMTP - Center for Applied Mathematics and Theoretical
  Physics, University of Maribor, Mladinska 3, SI-2000 Maribor, Slovenia,
  European Union}

\date{\today}

\begin{abstract}
We revisit the spectral statistics of the C$_3$-symmetric billiard introduced by Dembowski \textit{et al.} [Phys.\ Rev.\ E \textbf{62}, R4516 (2000)], which exhibits both GOE and GUE statistics depending on the symmetry block. Using high-precision Beyn’s contour-integral method for the nonlinear Fredholm eigenvalue problem with built-in separation of irreducible subspaces, we compute 2.8×10$^5$ eigenvalues in each symmetry subspace, enabling statistically meaningful comparisons with random matrix theory. The improved spectra reveal clear GOE--GUE correspondence and resolve previously observed deviations in long-range spectral correlations. Furthermore, we analyze phase--space eigenstate localization through the distribution of entropy localization measures, which, for chaotic states follow a Beta distribution whose standard deviation decays as a power--law with energy, consistent with the onset of quantum ergodicity as described by Schnirelman’s theorem.
\end{abstract}

\maketitle

\section{Introduction}
\label{introduction}

Quantum chaos (or more generally wave chaos) addresses the relationship between quantum or wave dynamics and the structure of the underlying classical phase space, particularly in the classical (short--wavelength) limit \cite{Stoe,Haake}. A central observation is how classical integrability, chaos, or mixed dynamics manifest themselves in quantum observables such as energy spectra, eigenfunctions, along with their phase--space representations. In this context, Wigner and Husimi functions \cite{Hus1940,Backer2003,voros1979,Wig1932} provide a natural framework for visualizing quantum states in phase space and identifying their correspondence with classical invariant structures. A fundamental result of quantum chaos is the universality of spectral statistics, where classically integrable systems exhibit uncorrelated energy levels with Poissonian statistics \cite{BerTab1977,Veble_Robnik_integrable}, whereas fully chaotic systems are described by Random Matrix Theory (RMT). Importantly, depending on the presence of antiunitary symmetries, the universality classes are described by either Gaussian Orthogonal (GOE) or Gaussian Unitary ensembles (GUE) \cite{robnik1986_GUE,robnik_proceedigs_1986_antiunitary}. This correspondence is known as the Bohigas-Giannoni-Schmit conjecture, which has its roots in the works of Casati, Guarnieri and Valz-Gris \cite{BGS1984,Cas1980}.

Billiard systems play a central role in studies of quantum chaos, as they provide conceptually simple yet dynamically rich models in which the correspondence between classical and quantum mechanics can be explored in detail. The classical dynamics of a billiard is fully determined by its geometry, ranging from integrable shapes (such as rectangles and circles) to fully chaotic ones (e.g., Sinai or Bunimovich billiards), as well as mixed-type geometries exhibiting both regular and chaotic motion \cite{Rob1983,Sinai1970,Bunimovich2001,Veble_Robnik_integrable,LLR2021,lozej2022triangles}. Their quantum counterparts, defined by the Helmholtz or Schrödinger equation with hard-wall boundary conditions, are experimentally accessible in microwave resonators, optical cavities, and acoustic systems, making billiards a natural testing ground for universal spectral statistics and phase–space localization phenomena {\cite{Stoe,Richter1998,experiment_microwave_tunneling,Jiang_laser,stockmann_microwave_billiards_2022}. For fully chaotic billiards, universality is not limited to spectral statistics but is also reflected in the structure of eigenfunctions. In the semiclassical limit, chaotic eigenstates are expected to become uniformly distributed in phase space, in accordance with the random-wave model \cite{berry1977}, whereas finite-energy effects give rise to deviations in the form of localization of eigenfunctions \cite{BLR2020,robnik_review_2023}, scars \cite{Heller1984}, and phenomenological level spacing distributions \cite{Bro1973,Izr1989,batistic_robnik_localization_eigenstates}. A quantitative study of eigenfunction localization is thus essential for determining how rapidly this universal behavior is attained.

Discrete symmetries of a billiard play a crucial role in shaping its quantum spectral statistics. The presence of a symmetry group leads to a decomposition of the Hilbert space into independent subspaces associated with the irreducible representations (irreps) of the group \cite{Irreps_ref}, and spectral statistics must therefore be analyzed separately within each irrep, as superpositions of levels from different symmetry sectors trivially drive the statistics towards Poissonian, therefore obscuring the physical origin of the statistical properties of the individual level sequences. In most billiard systems studied, the relevant symmetries are reflection symmetries, giving rise to one-dimensional irreps. In chaotic systems, each such symmetry-reduced spectrum follows the statistics of the Gaussian Orthogonal Ensemble, reflecting the presence of an effective time-reversal symmetry within each sector. In contrast, symmetry groups possessing higher-dimensional irreps can lead to a fundamentally different universality class. In such cases, the absence of an antiunitary symmetry within a given irrep results in Gaussian Unitary Ensemble statistics, even though the underlying classical dynamics remains time-reversal invariant \cite{c3_init}. Billiards with discrete rotational symmetries thus provide a natural but sparsely studied setting in which different irreps exhibit distinct random-matrix universality classes, offering a sensitive probe of the interplay between chaos, symmetry, and quantum statistics \cite{c4,dietz_threefold,dietz_c3_new}.

The paper is organized as follows. In Sec.~\ref{geometry}, we introduce the geometry of the billiard and discuss its symmetry properties. In Sec.~\ref{BIM}, we outline the boundary integral method used to solve the Helmholtz equation and to compute the eigenvalues and eigenfunctions. Section~\ref{spectral_statistics} is devoted to the analysis of spectral statistics, including nearest-neighbor level spacings and the spectral rigidity $\Delta_3$. In Sec.~\ref{A_measures}, we study the localization properties of eigenfunctions. In particular, Sec.~\ref{husimi} introduces the Poincar\'e--Husimi function, Sec.~\ref{localization_measures} its associated entropy localization measure, while Sec.~\ref{std_beta} analyzes the statistical behavior of the entropy localization measures by calculating the standard deviation of the fitted Beta distribution as a function of the wavenumber $k$. Sec.~\ref{conclusion_std_beta} analyzes and interprets the final results of Sec.~\ref{std_beta}. Finally, Sec.~\ref{general_conclusions} summarizes the results and presents our conclusions.

\section{Geometry}
\label{geometry}

In this study we have constructed the boundary from the following parametrization (Fig.\ref{fig:geometry_sketch}):

\begin{equation} \label{boundary_eq}
    r(\phi) = \tfrac{1}{2}\bigl(1 + a(\cos 3\phi - \sin 6\phi)\bigr) \quad \phi \in [0,2\pi) \quad a \in [0,0.55],
\end{equation}

\noindent where we primarily use $a=0.2$ \cite{c3_init,dietz_c3,dietz_c3_new} for the study of level spacings and entropy localization measures. Generally this billiard is of the mixed--type while being almost fully ergodic for $a \in [0.04,0.31]$ (Fig.\ref{fig:sali_grid}).

\begin{figure}
    \centering
    \includegraphics[width=1.0\linewidth]{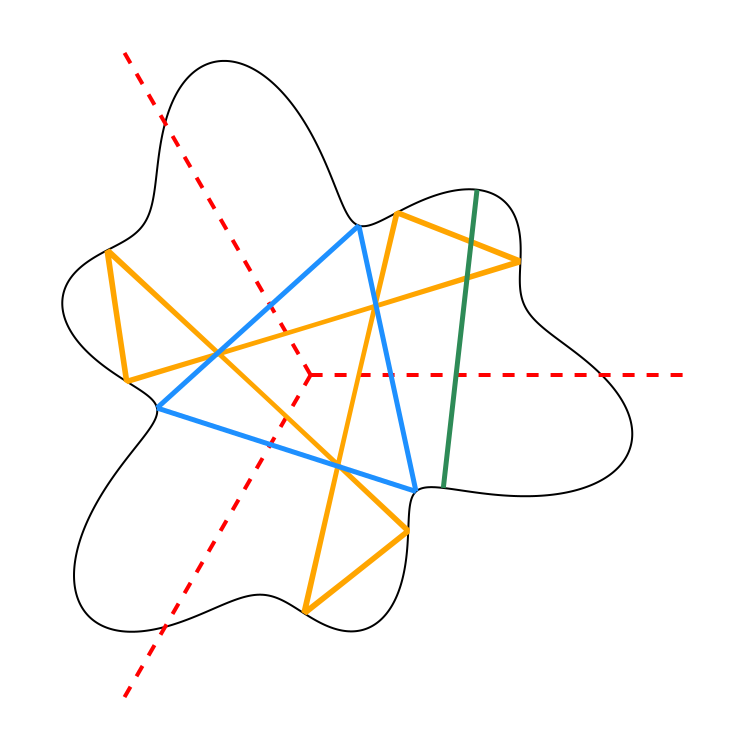}
    \caption{Primitive two- and three-bounce periodic orbits of the $C_3$ billiard at $a=0.2$. The boundary is shown in black, the three-bounce orbit in blue, the two-bounce orbit in green, and a six--bounce orbit in orange. Red dashed lines mark the symmetry axes defining the fundamental domain.}
    \label{fig:geometry_sketch}
\end{figure}

The C$_3$ billiard is invariant under the cyclic rotation group $C_3$ generated by
$R\equiv R_{2\pi/3}$ (rotation by $2\pi/3$). For $a\neq 0$ there are no reflections, so the
discrete symmetry is purely rotational. The Hilbert space decomposes into three one-dimensional
irreducible representations (irreps) labelled by the rotation eigenvalue
\begin{equation}
R\,\psi_m = e^{2\pi i m/3}\,\psi_m,\qquad m=0,1,2 .
\end{equation}
The $m=0$ sector is real (self-conjugate), while $m=1$ and $m=2$ are complex conjugates
and form a time-reversal pair (Table.\ref{tab:C3_irreps}).

\begin{table}[t]
\centering
\begin{tabular}{cccc}
\hline\hline
irrep & $m$ & eigenvalue of $R$ & time-reversal \\
\hline
$A$   & $0$ & $1$               & self-conjugate \\
$E_{+}$ & $1$ & $\omega=e^{+2\pi i/3}$ & $E_{-}$ \\
$E_{-}$ & $2$ & $\omega^2=e^{-2\pi i/3}$ & $E_{+}$ \\
\hline\hline
\end{tabular}
\caption{Irreducible representations of $C_3$ (rotation by $2\pi/3$), labeled as $R\equiv R_{2\pi/3}$.
The nontrivial eigenvalues are the cubic roots of unity $\omega=e^{2\pi i/3}$ and $\omega^2=e^{-2\pi i/3}$.}
\label{tab:C3_irreps}
\end{table}

\begin{figure}
    \centering
    \includegraphics[width=\linewidth]{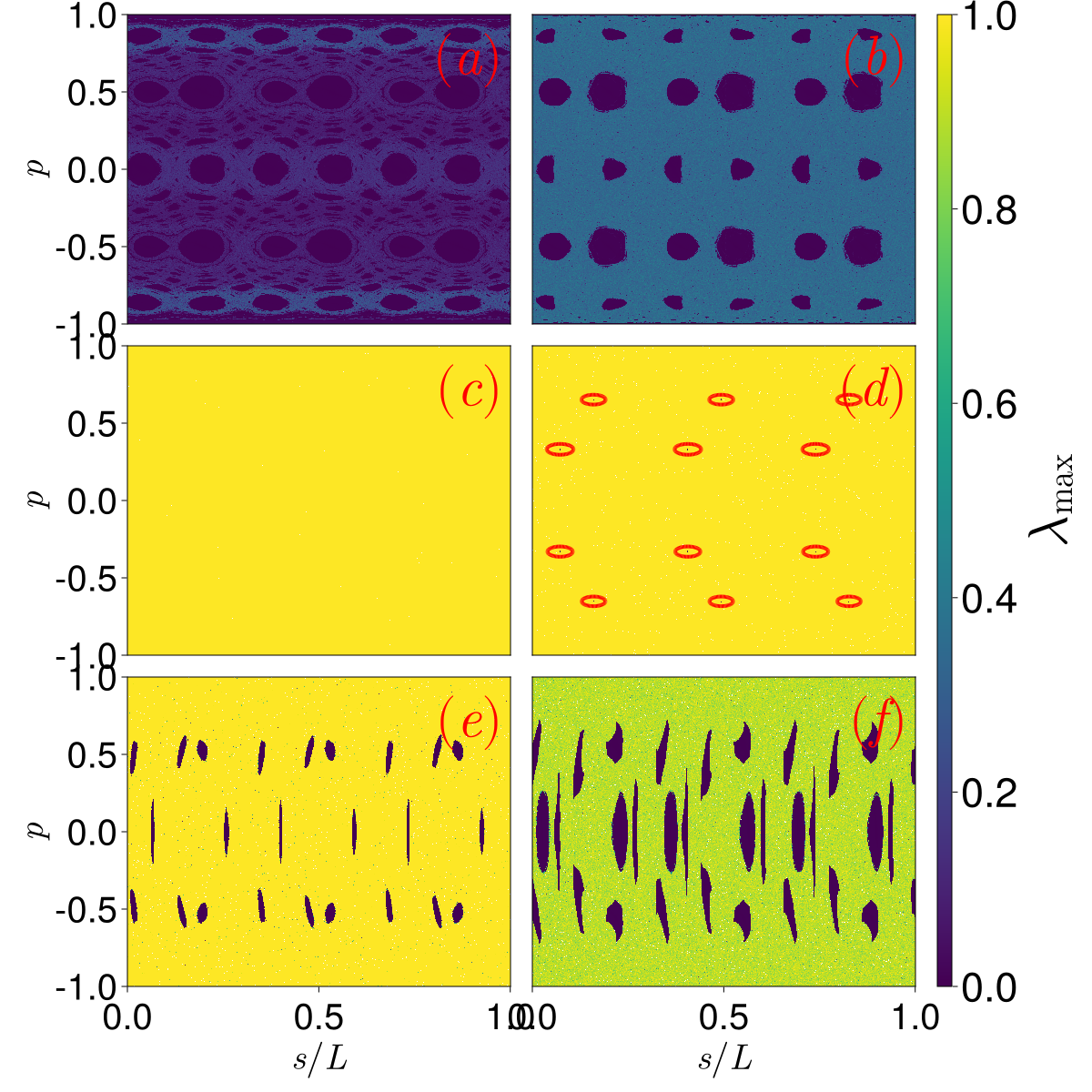}
    \caption{Finite-time maximal Lyapunov exponent heatmaps $\lambda_{\max}$ \cite{lyapunov_exponent} for the $C_3$ billiard on the Poincar\'e section in the fundamental domain $s/L\in[0,1/3)$. For visual clarity the color scale is capped at $\lambda_{\max}=1$ (values above the cap are shown with the same maximal color). Dark regions ($\lambda_{\max}\approx0$) indicate regular motion, whereas brighter regions correspond to chaotic dynamics with positive $\lambda_{\max}$. Panels show increasing deformation amplitude:(a) $a=0.005$, (b) $a=0.01$, (c) $a=0.05$, (d) $a=0.2$, (e) $a=0.35$, (f) $a=0.5$. For $0.05 \lesssim a \lesssim 0.31$, the phase space is predominantly chaotic; for larger $a$, it becomes mixed. Our case $a=0.2$ $(d)$ has very small stability islands (barely visible, encircled in $(d)$) corresponding to the orange PO in Fig.\ref{fig:geometry_sketch}. The heatmaps were constructed on a $1000 \times 1000$ grid with $5000$ collisions for each initial condition.}
    \label{fig:sali_grid}
\end{figure}

The finite-time maximal Lyapunov exponent is computed using the standard Benettin algorithm adapted to billiard systems. Between collisions, the
infinitesimal offset vector $\boldsymbol{\delta\Gamma}_m
=(\boldsymbol{\delta q}_m,\boldsymbol{\delta v}_m)$ evolves under the
linearized free-flight dynamics,

\begin{equation}
\begin{aligned}
\boldsymbol{q}_{m+1} = \boldsymbol{q}_m + \Delta t_m\ \cdot \boldsymbol{v}_{m}, \\
\boldsymbol{\delta v_{m+1}} = \boldsymbol{\delta v_m}
\\
\boldsymbol{\delta q_{m+1}} = \boldsymbol{\delta q_{m}} + \boldsymbol{\Delta t_m \delta v_m}
\end{aligned}
\label{eq:lyap_flight}
\end{equation}

\noindent where $\Delta t_m$ denotes the flight time between the $m$th and
$(m+1)$th collisions. At the $(m+1)$th collision with the boundary,
the $\delta\Gamma_{m+1}$ transforms according to the linearized reflection map (for details see \cite{lyapunov_billiards})

\begin{equation}
\begin{aligned}
\boldsymbol{\delta q}_{m+1}
&=
\boldsymbol{\delta q}_{m}-2\left( \boldsymbol{\delta q_m} \cdot \boldsymbol{n} \right)\boldsymbol{n}
\\
\boldsymbol{\delta v}_{m+1}
&=
\boldsymbol{\delta v}_m-2\left(\boldsymbol{\delta v}_m \cdot \boldsymbol{n}\right) \boldsymbol{n}- \frac{2\kappa}{\cos\alpha} \left( \boldsymbol{\delta q}_m \!\cdot\! \boldsymbol{t}_{i} \right) \boldsymbol{t}_{f},
\label{eq:lyap_collision}
\end{aligned}
\end{equation}

\noindent where $\boldsymbol{n}$ is the inward unit normal at the collision point, $\kappa$ the local boundary curvature at the collision point, $\alpha$ the incidence angle at the collision point, and $\boldsymbol{t}_{i}$, $\boldsymbol{t}_{f}$ denote the incoming and outgoing orthogonal unit vectors at the collision point (see Fig.2 for more information). To ensure numerical stability, the tangent-space vectors are periodically
orthonormalized using the QR algorithm \cite{benettin_normalization_1,benettin_normalization_2}. The finite-time maximal Lyapunov exponent \cite{lyapunov_exponent} is then obtained as

\begin{equation}
\lambda_{\max}(T)
=
\frac{1}{T}
\sum_{m=1}^{M(T)}
\ln
\frac{\|\boldsymbol{\delta\Gamma}_m\|}
{\|\boldsymbol{\delta\Gamma}_{m-1}\|},
\label{eq:lyap_definition}
\end{equation}

\noindent where $M$ is the total collision count and
$T=\sum_{m=1}^{M(T)} \Delta t_m$ is the total flight time.
Fig.~\ref{fig:sali_grid} shows a heatmap of maximal Lyapunov exponents
for characteristic deformation amplitudes $a$ in the boundary
definition~\eqref{boundary_eq}.

\section{Boundary Integral Method and the Helmholtz Equation}
\label{BIM}

We consider the stationary Helmholtz equation
\begin{equation}
(\Delta + k^2)\psi(\mathbf{x}) = 0, \qquad \mathbf{x} \in \Omega,
\label{eq:helmholtz}
\end{equation}
in a two-dimensional billiard domain $\Omega$ with Dirichlet boundary conditions
\begin{equation}
\psi(\mathbf{x}) = 0,
\qquad \mathbf{x} \in \partial\Omega,
\label{eq:dirichlet_bc}
\end{equation}

\noindent where $\partial \Omega$ is the total boundary parametrized as in \eqref{boundary_eq}. For simply connected domains with Dirichlet boundary conditions, we employ a boundary integral formulation based on the double-layer potential \cite{backer_BIM} where 

\begin{equation} \label{boundary_function}
    u(\mathbf{y})=\partial_{\mathbf{n}_{\mathbf{y}}}(\psi(\mathbf{y}))
\end{equation}

\noindent is the boundary function, $\partial_{\mathbf{n}_{\mathbf{y}}}$ denoting the outward normal derivative at $\mathbf{y}$, and

\begin{equation}
G_k(\mathbf{x},\mathbf{y})
=-
\tfrac{i}{4}
H_0^{(1)}\!\left(k|\mathbf{x}-\mathbf{y}|\right)
\label{eq:green_function}
\end{equation}

\noindent being the free-space Green's function and $H_{m}^{(1)}$ being the Hankel function of the first kind of order $m$ \cite{abramowitz}. The boundary integral equation (BIE) is

\begin{equation} \label{eq:dlp_bie}
A(u(\mathbf{x})) =\tfrac{1}{2}\,u(\mathbf{x})
+ 
\int_{\partial \Omega}
\partial_{n_{\mathbf{y}}} G_k(\mathbf{x},\mathbf{y})\,
u(\mathbf{y})\,ds(\mathbf{y})
=0,
\qquad \mathbf{x}\in \partial \Omega
\end{equation}

\noindent where the $\frac{1}{2}u(\mathbf{x})$ term follows from the jump relation of the double-layer operator \cite{kress,backer_BIM} and $\partial_{n_{\mathbf{y}}} G_{{k}}(\mathbf{x},\mathbf{y})$ is given as

\begin{equation} \label{eq:DLP_kernel}
    \partial_{n_{\mathbf{y}}} G_{k}(\mathbf{x},\mathbf{y}) = - \frac{ik}{4}\frac{\mathbf{n(y)}\cdot(\mathbf{x}-\mathbf{y})}{d(\mathbf{x},\mathbf{y})}H_{1}^{(1)}(kd(\mathbf{x},\mathbf{y}))
\end{equation}

\noindent where $d(\mathbf{x},\mathbf{y})=\sqrt{\sum_{i=1}^{N}(x_i-y_i)^2}$ is the Euclidian distance between the points and $\mathbf{n(y)}$ the outward pointing normal vector at point $\mathbf{y}$. Due to the present $C_3$ rotational symmetry symmetry it is numerically easier to work with the fundamental domain $\Gamma_{fund}$ instead of the full domain $\Gamma_{full}$. They are defined as

\begin{equation}
\Gamma_{\mathrm{fund}} = \Gamma\!\left(\tfrac{2\pi}{3}\right),
\qquad
\Gamma_{\mathrm{full}} = \Gamma(2\pi) = \partial \Omega,
\label{eq:gammas}
\end{equation}

\noindent where
\[
\Gamma(\theta)
=
\left\{
\bigl(r(\phi)\cos\phi,\; r(\phi)\sin\phi\bigr)
\;\middle|\;
\phi\in[0,\theta)
\right\}.
\]

\noindent The BIE \eqref{eq:dlp_bie}  can be reduced to a boundary integral over $\Gamma_{fund}$ and decomposed into irreducible symmetry sectors labeled by $m=0,1,2$. The resulting symmetry-projected kernel \cite{dietz_c3_new} is (for derivation see Appendix.\ref{irrep_bie_derivation})

\begin{equation} \label{eq:bie_desymm_main_text}
\begin{aligned}
    A(u_m(\mathbf{x})) = \frac{1}{2}u_m(\mathbf{x}) +  \sum_{\ell=0}^{2}\chi_m(\mathbf{R}^{\ell})^{*} \int_{\Gamma_{\mathrm{fund}}} \partial_{n_{\mathbf{y}}} G_{k}(\mathbf{x},\mathbf{y}),u_m(\mathbf{y})\,ds(\mathbf{y}),
\end{aligned}
\end{equation}

\noindent where $u_m(\mathbf{x})$ is the boundary function in the $m$-th irreducible representation, $\mathbf{R}$ the rotation operator given in \eqref{eq:rotation_operator} and  $\chi_m(\mathbf{R^\ell})^*$ being the complex conjugate of the character of the rotation $\mathbf{R^\ell}$ in the $m$-th irreducible representation (see \eqref{eq:character_exp}). Similarly the wavefunction constructed using \eqref{eq:green_function}

\begin{equation}
\psi(\mathbf{x})
=
-
\int_\Gamma
G_k(\mathbf{x},\mathbf{y})\,
\varphi(\mathbf{y})\,ds(\mathbf{y}),
\qquad \mathbf{x}\in\Omega,
\label{eq:dlp_representation}
\end{equation}

\noindent can be constructed in a given irreducible representation $m$ (for derivation see Appendix.\ref{irrep_wavefunction_derivation})

\begin{equation} \label{eq:psi_slp_fund_main}
\psi_m(\mathbf{x})
=
\sum_{\ell=0}^{2}\chi_m(\mathbf{R}^{\ell})^{*}
\int_{\Gamma_{\mathrm{fund}}}
G_k(\mathbf{x},\mathbf{R}^{\ell}\mathbf{y})\,
u_m(\mathbf{y})\,ds(\mathbf{y}),
\qquad \mathbf{x}\in\Omega.
\end{equation}

\section{Spectral statistics}
\label{spectral_statistics}

\subsection{Nearest neighbor level spacings - NNLS}

\begin{figure}
    \centering

    \begin{overpic}[width=\linewidth]{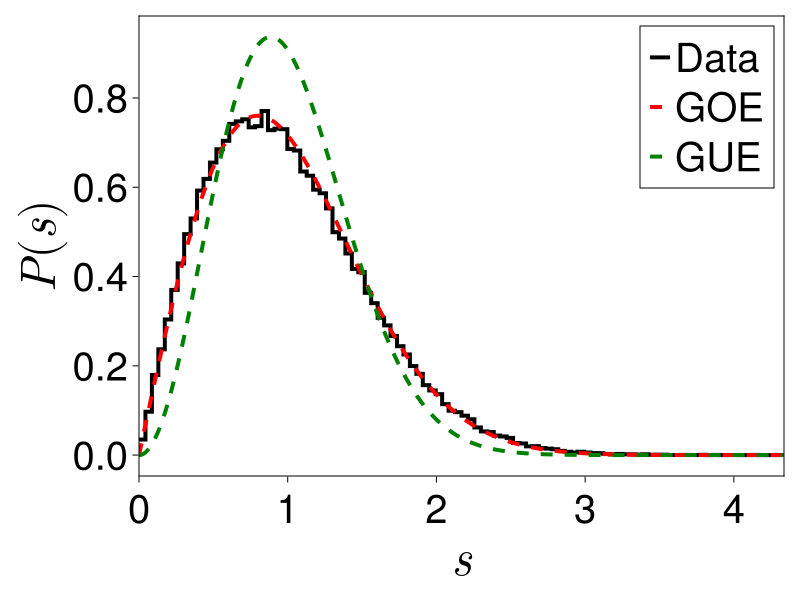}
        \put(20,65){\large\bfseries (a)}
    \end{overpic}
    \hfill
    \begin{overpic}[width=\linewidth]{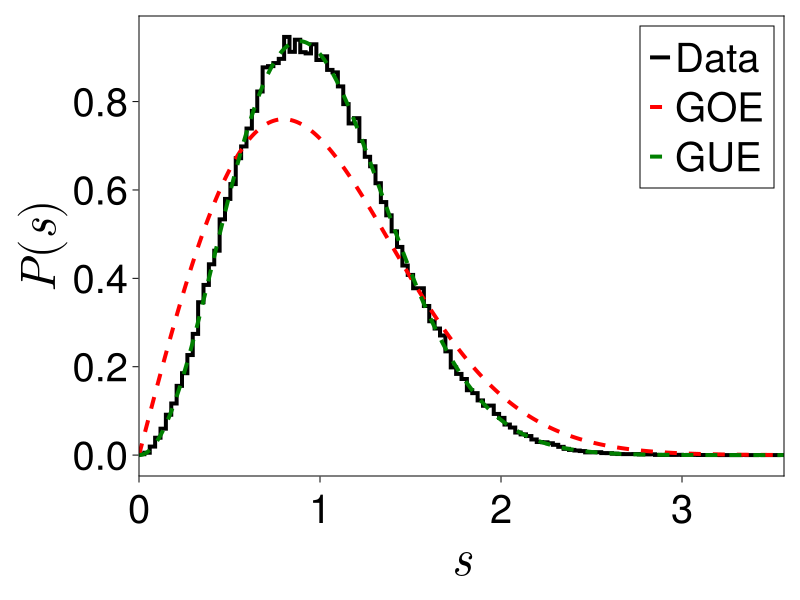}
        \put(20,65){\large\bfseries (b)}
    \end{overpic}

    \caption{
    Nearest-neighbor level spacing distributions for the desymmetrized spectrum of the
    $C_3$ billiard.
    (a) $m=0$ symmetry sector: the numerical histogram (approximately $2.8\times10^5$
    unfolded eigenvalues) is compared with the Wigner surmise of the Gaussian Orthogonal
    Ensemble (GOE), reflecting the presence of time-reversal symmetry in this real
    irreducible representation.
    (b) $m=1$ symmetry sector: comparison with the Gaussian Unitary Ensemble (GUE)
    Wigner surmise, indicating effective time-reversal symmetry breaking in the complex
    irreducible representations.
    }
    \label{fig:nnls}
\end{figure}

To characterize short-range spectral correlations, we analyze the nearest-neighbor
level spacing distribution (NNLS) of the unfolded spectrum. After unfolding, spacings
are defined as
\begin{equation}
s_n = E_{n+1} - E_n,
\end{equation}
where $\{E_n\}$ denotes the unfolded eigenvalues ordered by increasing energy. Owing to the exact $C_3$ rotational symmetry of the billiard, the spectrum decomposes into statistically independent symmetry sectors corresponding to the irreducible representations $m=0,1,2$. Since the $m=1$ and $m=2$ sectors form a complex-conjugate pair, having the same energy (doublets), related by time-reversal symmetry, we restrict the analysis to the $m=0$ and $m=1$ sectors. The NNLS distributions for the desymmetrized spectra in the $m=0$ and $m=1$ symmetry sectors are shown in Fig.~\ref{fig:nnls}, each based on approximately $2.8\times10^5$ eigenvalues.

The numerical results are compared with the Wigner surmises
\begin{align}
P_{\mathrm{GOE}}(s) &= \frac{\pi}{2}\, s\, \exp\!\left(-\frac{\pi s^2}{4}\right), \\
P_{\mathrm{GUE}}(s) &= \frac{32}{\pi^2}\, s^2\, \exp\!\left(-\frac{4 s^2}{\pi}\right),
\end{align}
corresponding to the Gaussian Orthogonal Ensemble (GOE) and Gaussian Unitary Ensemble
(GUE), respectively.

We observe excellent agreement with GOE statistics in the $m=0$ sector, consistent
with the presence of time-reversal symmetry in this real irrep. In contrast, the
$m=1$ sector follows GUE statistics, reflecting the effective breaking of
time-reversal symmetry in the complex irreps. This behavior is expected for systems with discrete rotational symmetry, where complex irreps support unitary symmetry classes leading to GUE statistics even in the absence of an external
magnetic field \cite{dietz_poly_basis_method_c3}.

\subsection{Spectral rigidity - \texorpdfstring{$\Delta_3$}{Delta3}}

While nearest-neighbor spacing statistics probe short-range correlations, long-range
spectral correlations are characterized by the Dyson--Mehta \cite{Mehta} spectral rigidity $\Delta_3(L)$. For an unfolded spectrum $\{E_n\}$ with unit mean level density, we introduce the spectral staircase function

\begin{equation}
N(E)=\#\{n:\,E_n \le E\}.
\label{eq:staircase}
\end{equation}

\noindent The rigidity measures the mean-square deviation of $N(E)$ from its best linear fit on
an interval of length $L$,

\begin{equation}
\Delta_3(L)
=
\left\langle
\min_{A,B}
\frac{1}{L}
\int_{E_0}^{E_0+L}
\left[
N(E)-(A E + B)
\right]^2
\, dE
\right\rangle_{E_0},
\label{eq:delta3_def}
\end{equation}

\noindent where $\langle\cdots\rangle_{E_0}$ denotes averaging over the energy interval $[E_0,E_0 + L]$ . An equivalent characterization of long-range correlations is provided by the number variance

\begin{equation}
\Sigma^2(L)
=
\left\langle
\left[
N(E_0+L)-N(E_0)-L
\right]^2
\right\rangle_{E_0},
\label{eq:number_variance}
\end{equation}

\noindent which measures fluctuations of the number of levels in an interval of length $L$. The two quantities are related by the identity \textcolor{red}{\cite{Stoe}}

\begin{equation}
\Delta_3(L)
=
\frac{2}{L^4}
\int_0^L
\left(
L^3 - 2L^2 r + r^3
\right)
\Sigma^2(r)
\, dr.
\label{eq:delta3_sigma2_relation}
\end{equation}

\begin{figure}
    \centering
    \includegraphics[width=\linewidth]{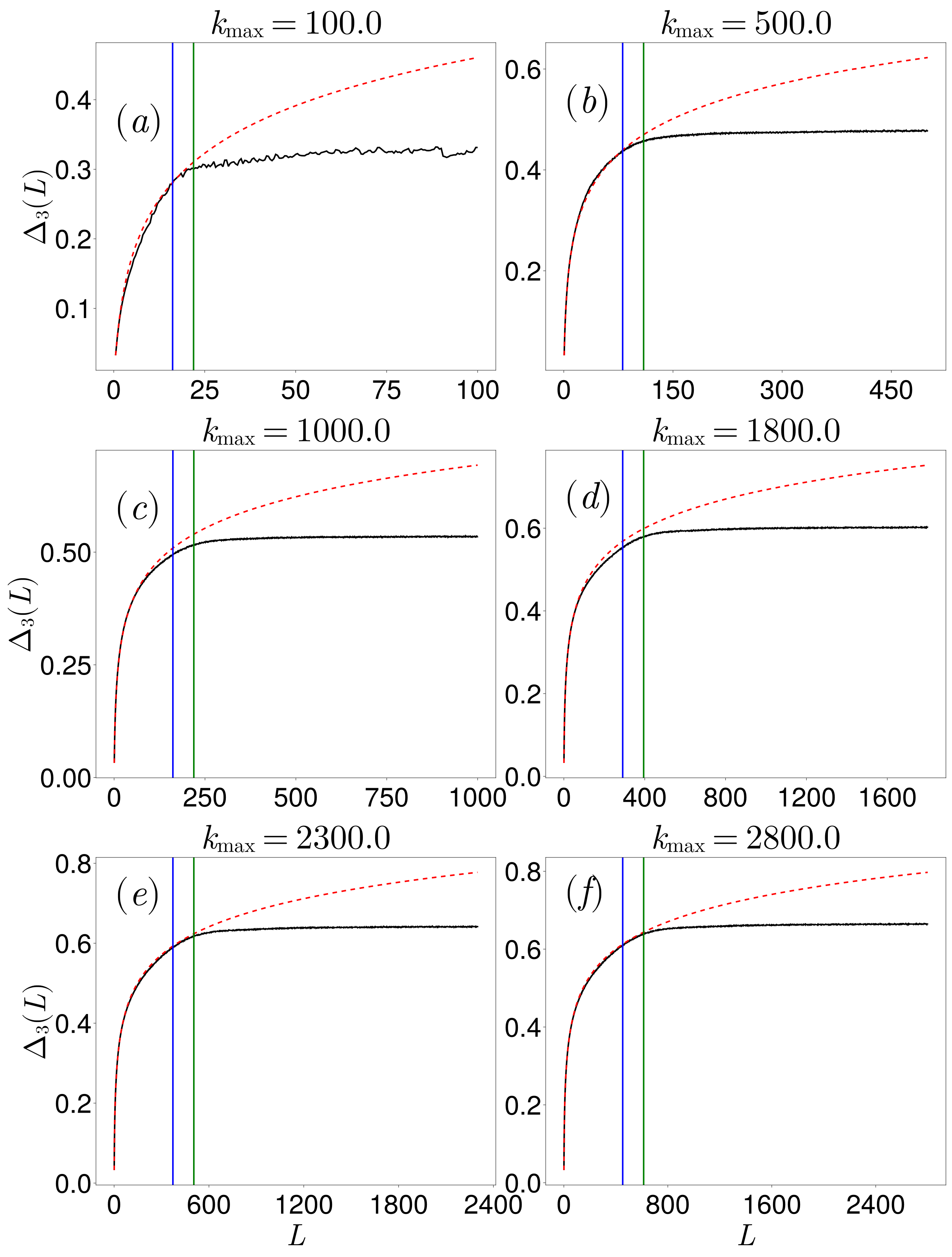}
    \caption{$\Delta_3(L)$ for the $m=0$ irreducible representation of the  billiard with $a=0.2$ at various wavenumbers $k$. The dashed red line is the theoretical curve for the $\Delta_3^{GOE}(L)$, the blue vertical represents the length of the 3--bounce periodic orbit and the green line represents the 2--bounce periodic orbit, as shown in Fig.\ref{fig:geometry_sketch}.It is observed that the spectrum starts to deviate from RMT around $L_{max} \approx\frac{A_{fund} \times k_{max}}{l_0}$, which matches the value of $l_0 = 1.2415$ corresponding to the $2$--bounce orbit (green vertical line, shown in green in Fig.\ref{fig:geometry_sketch}), where $A_{fund}$ is the area of the desymmetrized domain ($A_{fund} = \frac{A}{3}$) and $k_{max}$ the largest wavenumber present in the spectrum (title in each subfigure) \cite{Berry1985,prosen_shortest_PO}. The blue vertical line corresponding to the $3$--bounce orbit with $l_0 = 1.6834$ (shown in blue in Fig.\ref{fig:geometry_sketch}) which was used in \cite{dietz_c3_new} does not correctly capture the observed saturation trend.}
    \label{fig:delta_prog_0}
\end{figure}

For uncorrelated (Poissonian) spectra, the rigidity grows linearly,

\begin{equation}
\Delta_3^{\mathrm{Poisson}}(L)=\frac{L}{15},
\label{eq:delta3_poisson}
\end{equation}

\noindent whereas for chaotic systems random-matrix theory predicts a much slower, logarithmic
growth. In the Gaussian ensembles, the asymptotic behavior at large $L$ reads

\begin{figure}
    \centering
    \includegraphics[width=\linewidth]{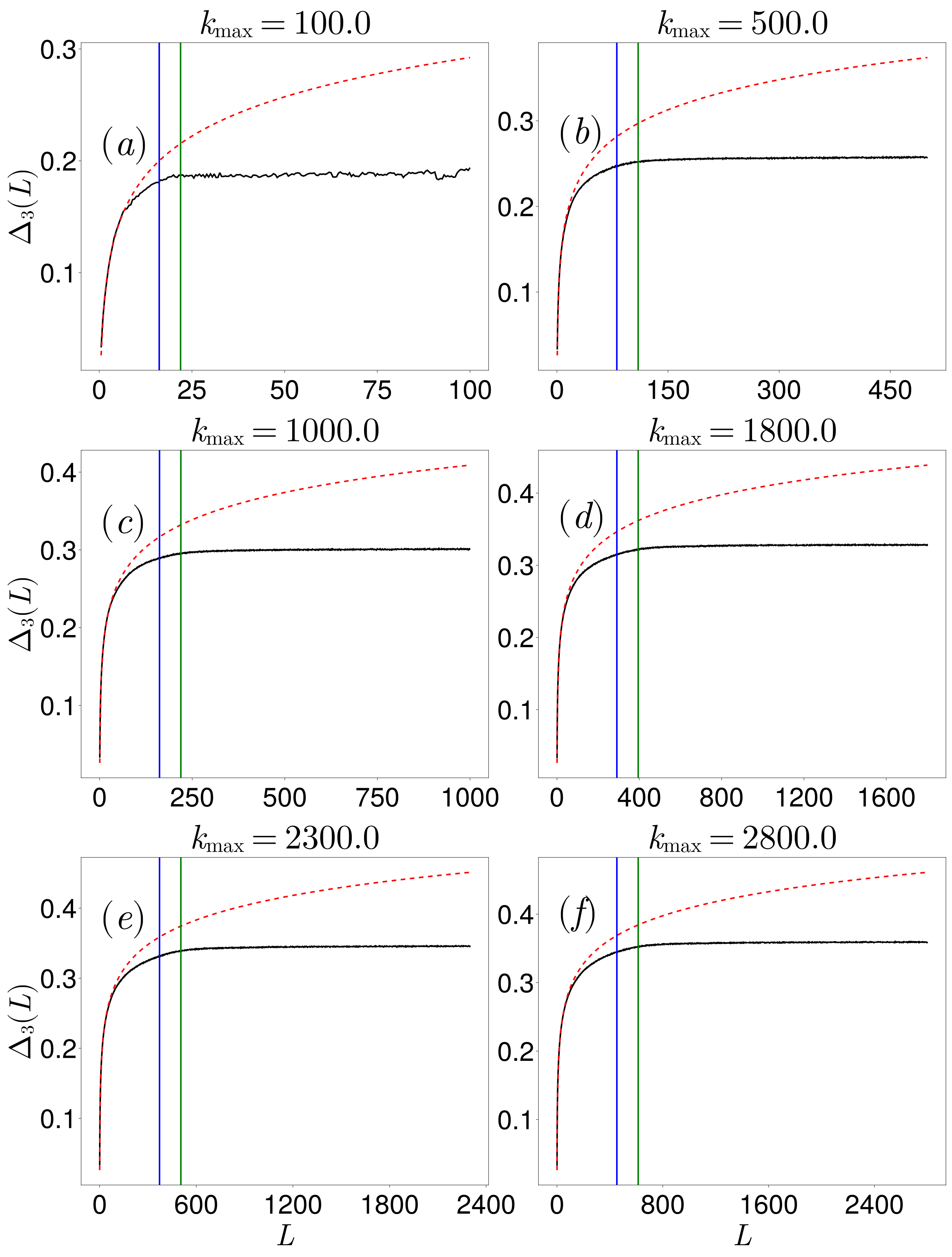}
    \caption{$\Delta_3(L)$ for the $m=1$ irreducible representation of the  billiard with $a=0.2$ at various wavenumbers $k$. The plotted quantities are the same as in Fig.\ref{fig:delta_prog_0}, where importantly the value of $L_{max}$ is irrep agnostic, allowing for a direct comparison between saturation trend. In contrast to the $m=0$ case the deviation from $\Delta_3^{GUE}(L)$ RMT is observed already before the prediction given by the length of the shortest periodic orbit (green vertical line), while the saturation is still observed to start at the calculated $L_{max}$.}
    \label{fig:delta_prog_1}
\end{figure}

\begin{equation}
\Delta_3^{(\beta)}(L)
\sim
\frac{1}{\beta\pi^2}\,\ln L + C_\beta,
\qquad
\beta=
\begin{cases}
1, & \text{GOE},\\
2, & \text{GUE},
\end{cases}
\label{eq:delta3_rmt_asympt}
\end{equation}

\noindent with ensemble-dependent constants $C_\beta$. For the case of $a=0.2$, $\Delta_3(L)$ is shown for symmetry classes $m=0,1$ in Fig.\ref{fig:delta_prog_0} and Fig.\ref{fig:delta_prog_1}, respectively. It shows agreement with spectral rigidity RMT up until saturation due to the shortest periodic orbit.
Another complementary result is the dependence of the saturation value of the $\Delta_3^{\infty}$ as a function of wavenumber $k$. The saturation value represents the maximal possible value of spectral rigidity achievable at a given maximal energy $E_{max}$ or equivalent wavenumber $k_{max}$ in the spectrum and signifies the contribution of the short classical periodic orbit. As described in detail in \cite{Berry1985}, the growth of $\Delta_3^{\infty}$ is logarithmic in $E=k^2$ and universal in each symmetry class, while the additive constant $C$ is system-dependent. The trend is clearly observed in Fig.\ref{fig:m_01_saturation}, showing a clear agreement of the numerical data with the theoretical result.

\begin{figure}
    \centering
    \begin{overpic}[width=\linewidth]{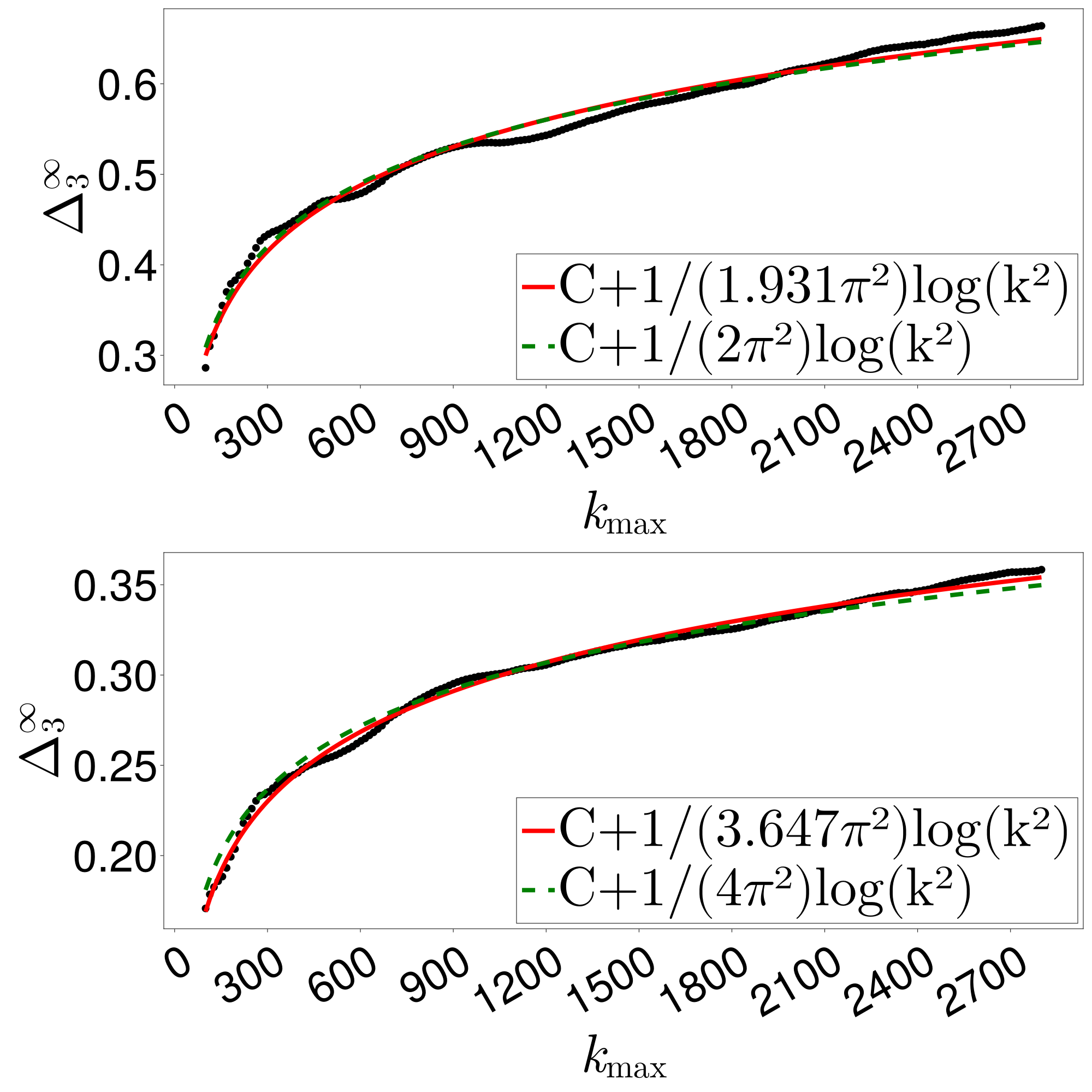}
    \put(25,90){\large a)}
    \put(25,40){\large b)}
    \end{overpic}
    \caption{Spectral rigidity $\Delta_{3}^{\infty}$ as a function of the maximal wavenumber $k_{\max}$ for the $m=0$ (GOE, top panel) and $m=1$ (GUE, bottom panel) symmetry classes. Black dots denote numerical results, while the solid red curves show fits of the form $\Delta_{3}^{\infty}$=$C+\frac{1}{\alpha\pi^2}\ln(k^2)$ with $\alpha=1.931$ (GOE) and $\alpha=3.647$ (GUE). The value of $C$ is billiard dependant and was determined by fitting. Theoretical values are represented by the dashed green line showing agreement with the numerical results.}
    \label{fig:m_01_saturation}
\end{figure}

\section{Phase--space entropy localization measures}
\label{A_measures}

\subsection{Poincaré--Husimi function}
\label{husimi}

To analyze the phase--space structure of individual eigenstates we employ Poincaré--Husimi (PH) functions \cite{voros1979}, which provide a reduced phase--space representation on the billiard boundary. This construction is particularly well suited for billiard systems, where the boundary acts as a natural global
Poincaré section \cite{Steiner1994}.

\begin{figure}
\centering
\begin{overpic}[width=\linewidth]{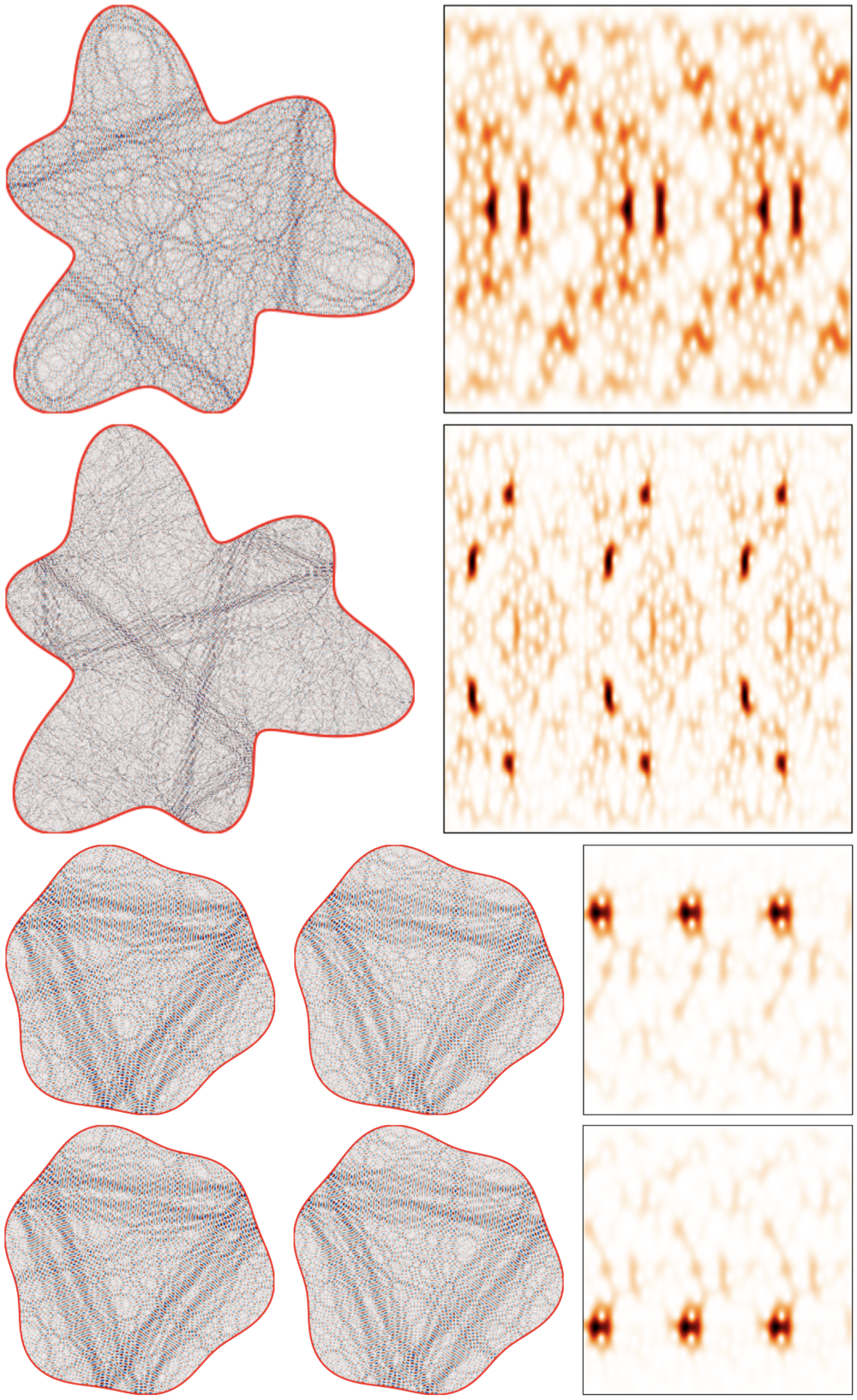}
  \put(25,95){\large a)}
  \put(25,65){\large b)}
  \put(19,38){\large c)}
  \put(19,18){\large d)}
\end{overpic}
\caption{$a)$ Probability density of a scar associated with the shortest periodic orbit (two-bounce orbit) together with the corresponding PH function for $a=0.2$ and $m=0$ . $b)$ Probability density with a scar like feature supported by the periodic orbit linked to the stability islands visible in the finite-time maximal Lyapunov exponent phase-space analysis, along with its PH representation for $a=0.2$ and $m=0$ (see Figs.\ref{fig:geometry_sketch},\ref{fig:sali_grid}(d)). $c)$ Real and imaginary parts of a quantum eigenfunction with $a=0.05$, $m=1$ and $k \approx 331.25$. While the individual components do not exhibit explicit $2\pi/3$ rotational symmetry, their probability density preserves the symmetry. The corresponding PH function displays the breaking of the $p \rightarrow -p$ symmetry. d) shows the doublet partner of c) with $a=0.05$, $m=2$ and $k \approx 331.25$. For the relevant periodic orbits seen in a) and b) see Fig.\ref{fig:geometry_sketch}.}
\label{fig:PO_and_symmetry_break_husimi}
\end{figure}
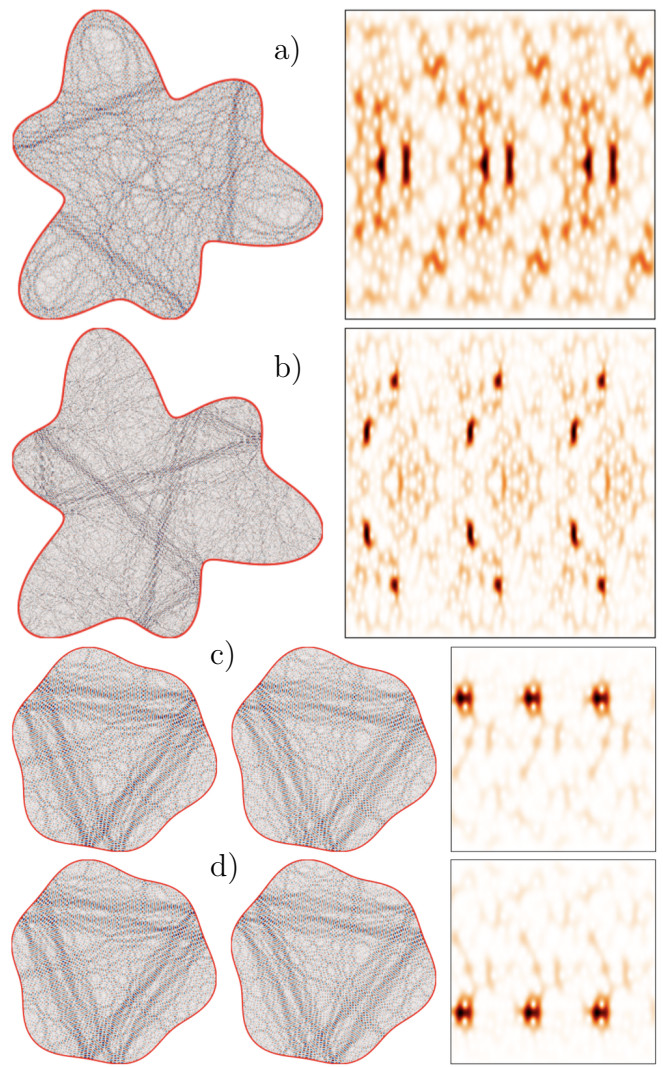

Let $\psi_n(\mathbf{x})$ be an eigenfunction satisfying \eqref{eq:helmholtz}. Then the eigenfunction is uniquely determined by its normal derivative on the boundary \cite{backer_boundary}, which is the boundary function $u_n(s)$ defined in \eqref{boundary_function}. The Poincaré section is parameterized by boundary position $q\in[0,L)$
and momentum

\begin{equation}
p = \sin(\alpha),
\end{equation}

\noindent where $\alpha$ is the reflection angle of the specular reflection.

Following Ref.~\cite{Backer2003}, coherent states on the boundary are defined as
periodized Gaussian wave packets,

\begin{equation}
\begin{aligned}
c_{(q,p),k}(s)
&=
\left(\frac{k}{\pi}\right)^{1/4}
\sum_{m\in\mathbb{Z}} \times \\
&\quad
\exp\!\left[
-\frac{k}{2}(s-q+mL)^2
- i k p (s-q+mL)
\right].
\end{aligned}
\end{equation}

\noindent and the unnormalized Poincaré--Husimi function associated with the eigenstate $\psi_n$ is defined as
\begin{equation}
h_n(q,p)
=
\left|
\int_0^{L}
c_{(q,p),k_n}(s)\,u_n(s)\,ds
\right|^2.
\label{eq:ph_def}
\end{equation}

\noindent This quantity represents the projection of the boundary function $u_n(s)$
onto a minimum--uncertainty wave packet centered at $(q,p)$ on the Poincaré section. Due to the fact that we are working in irreps the PH functions were constructed with symmetry reduction in mind, allowing for less numerical effort in their evaluation (for derivation see Appendix.\ref{irrep_ph_derivation}). Examples are shown in Fig.\ref{fig:PO_and_symmetry_break_husimi}(a,b,c,d).

\begin{figure}
    \centering
    \includegraphics[width=\linewidth]{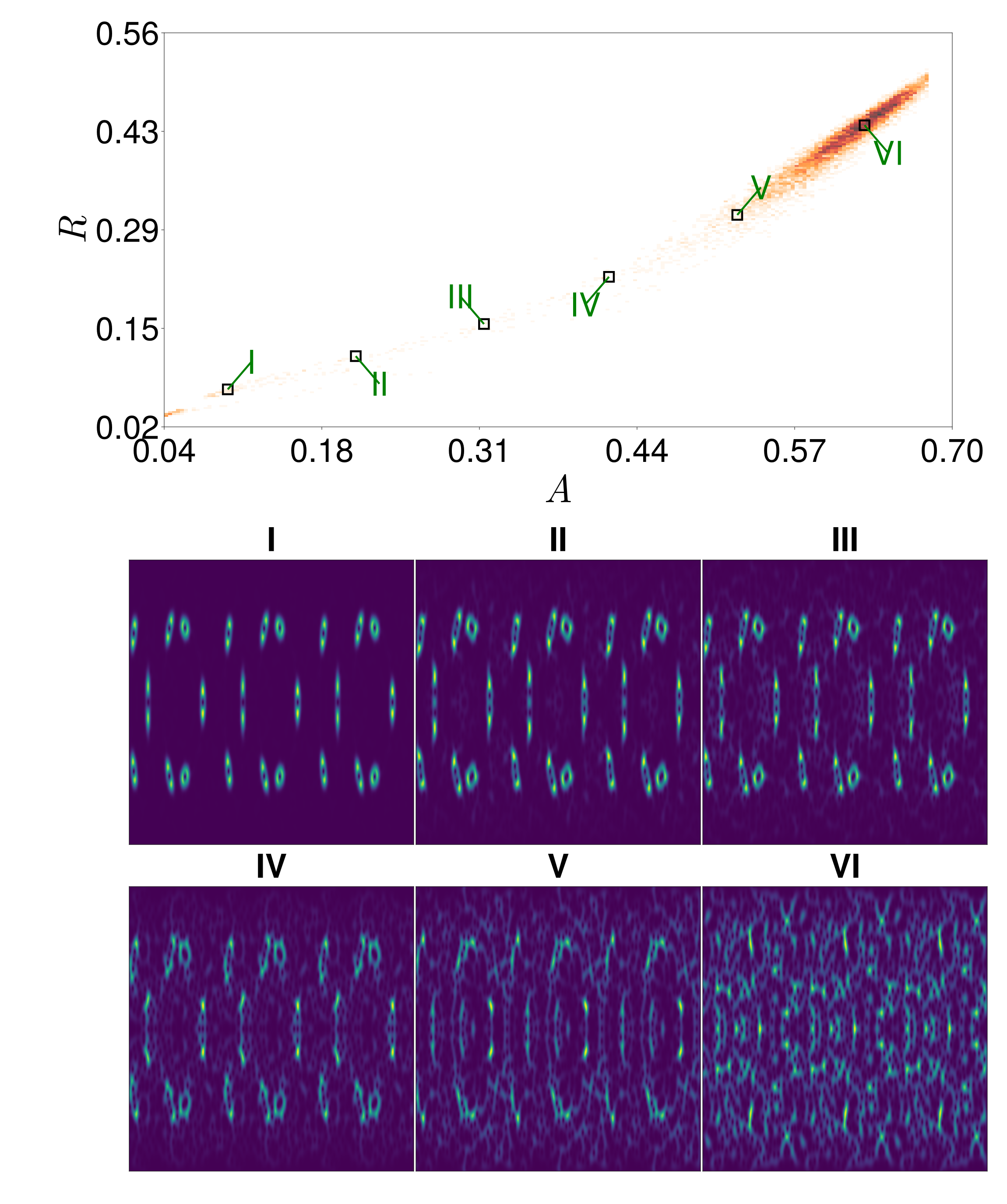}
    \caption{Localization measures for the C$_3$ billiard with $a=0.35$ in the $m=0$ irreducible subspace from $k=800$ to $k=1000$ (approx. $8500$ eigenstates). Top: joint distribution of the localization entropy $A$ and the normalized inverse participation ratio $R$, defined as $R=\frac{1}{N_c}\frac{1}{\sum_{i,j}{H_{i,j}^2}}$, where $N_c$ is the number of chaotic grid cells. Highlighted states (I–VI) are representative examples. Bottom: corresponding PH in the full domain (see Fig.\ref{fig:sali_grid}((e)) for the classical phase space). Brighter spots correspond to larger PH function values while dark blue corresponds to $0$. The distribution clearly separates into a dominant bulk and sparse tails. The bulk corresponds to chaotic eigenstates, centered around $A_{max} \approx 0.66$ and well described by a fitted Beta distribution \cite{BLR2019B,LLR2021,BLR2020}. In contrast, the tails at low $A$ originate from regular or residual mixed-type states, characterized by strong phase-space localization. For the purpose of statistical modeling of chaotic eigenstates, these tail states are excluded from the fitting procedure \cite{BatManRob2013}. This filtering isolates the chaotic ensemble and yields stable Beta distribution parameters, while the removed states represent physically distinct dynamical classes.}
    \label{fig:ipr_a_0.34_m_0}
\end{figure}

The symmetry of $h_n(q,p)$ reflects the underlying discrete symmetries
of the billiard.
In particular:

\begin{figure}
    \centering
    \includegraphics[width=\linewidth]{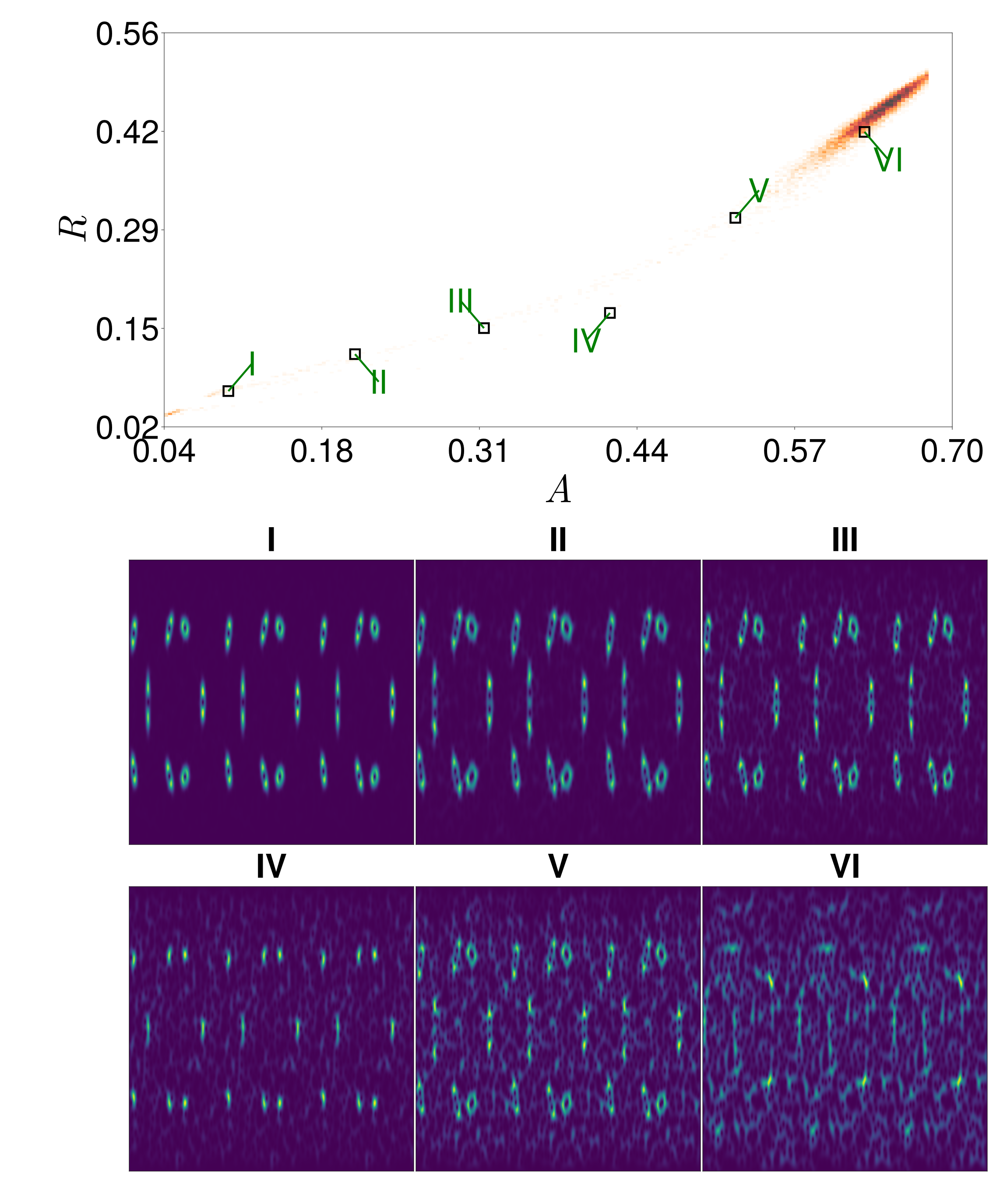}
    \caption{Localization measures for the C$_3$ billiard with $a=0.35$ in the $m=1$ irreducible subspace from $k=800$ to $k=1000$. The detailed description mirrors Fig.\ref{fig:ipr_a_0.34_m_0}. It is observed that comparing the width of the fitted Beta distribution between $m=0$ and $m=1$ irreducible subspaces shows consistent narrowing in the $m=1$ case.}
    \label{fig:ipr_a_0.34_m_1}
\end{figure}

\begin{itemize}
\item For real irreducible representations (time--reversal invariant,
GOE--like sectors, $m=0$, Figs.\ref{fig:PO_and_symmetry_break_husimi}(a,b)),

\begin{equation}
h_n(q,p)=h_n(q,-p).
\end{equation}

\item For complex irreducible representations (GUE--like sectors, m=1,2),
time--reversal symmetry is broken and no such $p\to -p$ symmetry exists (Fig.\ref{fig:PO_and_symmetry_break_husimi}(c,d)).
\end{itemize}

\subsection{Entropy of localization measures}
\label{localization_measures}

To quantify the degree of phase--space localization,
we discretize $h_n(q,p)$ on a uniform $(q,p)$ grid
and normalize it such that $\sum_{q,p} h_n(q,p)=1$ \cite{BLR2018}.
The localization entropy is then defined as

\begin{equation}
S_n
=
-\sum_{q,p} h_n(q,p)\,\log h_n(q,p),
\end{equation}

\noindent and the corresponding normalized localization measure,

\begin{equation}
A_n = \frac{e^{S_n}}{N_{\mathrm{eff}}},
\label{eq:An_def}
\end{equation}

\noindent where $N_{\mathrm{eff}}$ denotes the number of accessible chaotic phase--space cells. By construction, $A_n\in(0,1)$: small values of $A_n$ indicate strong localization (e.g.\ scarring or regular islands), while values close to $0.66$ correspond to fully extended states in phase--space \cite{Yan_FPUT_mixed}. The distribution $P(A)$ of localization measures
is generally asymmetric and broad. Empirically it is found that $P(A)$ for localized chaotic states is well described by a Beta distribution \cite{BLR2020},

\begin{equation} \label{beta_distr}
P(A;\alpha,\beta)
=
C A^{\alpha-1}(A_0-A)^{\beta-1}
\qquad 0<A<A_0,
\end{equation}

\noindent with normalization $C^{-1}=A_0^{\alpha+\beta - 1} B(\alpha,\beta)$ and $B(\alpha,\beta)=\int_{0}^{1}t^{\alpha-1}(1-t)^{\beta-1}$ is the Euler Beta function. The shape parameters $(\alpha,\beta)$ encode the relative weight of localized versus extended eigenstates and vary systematically with energy and classical phase--space structure. Examples for the case of $a=0.2$ with $m=0,1$ symmetry sectors are given in Fig.\ref{fig:PA_GOE} and Fig.\ref{fig:PA_GUE}. Provided are also Fig.\ref{fig:ipr_a_0.34_m_0} and Fig.\ref{fig:ipr_a_0.34_m_1} as cases of a mixed--type phase space with localization of various degrees.

\begin{figure}
    \centering
    \includegraphics[width=\linewidth]{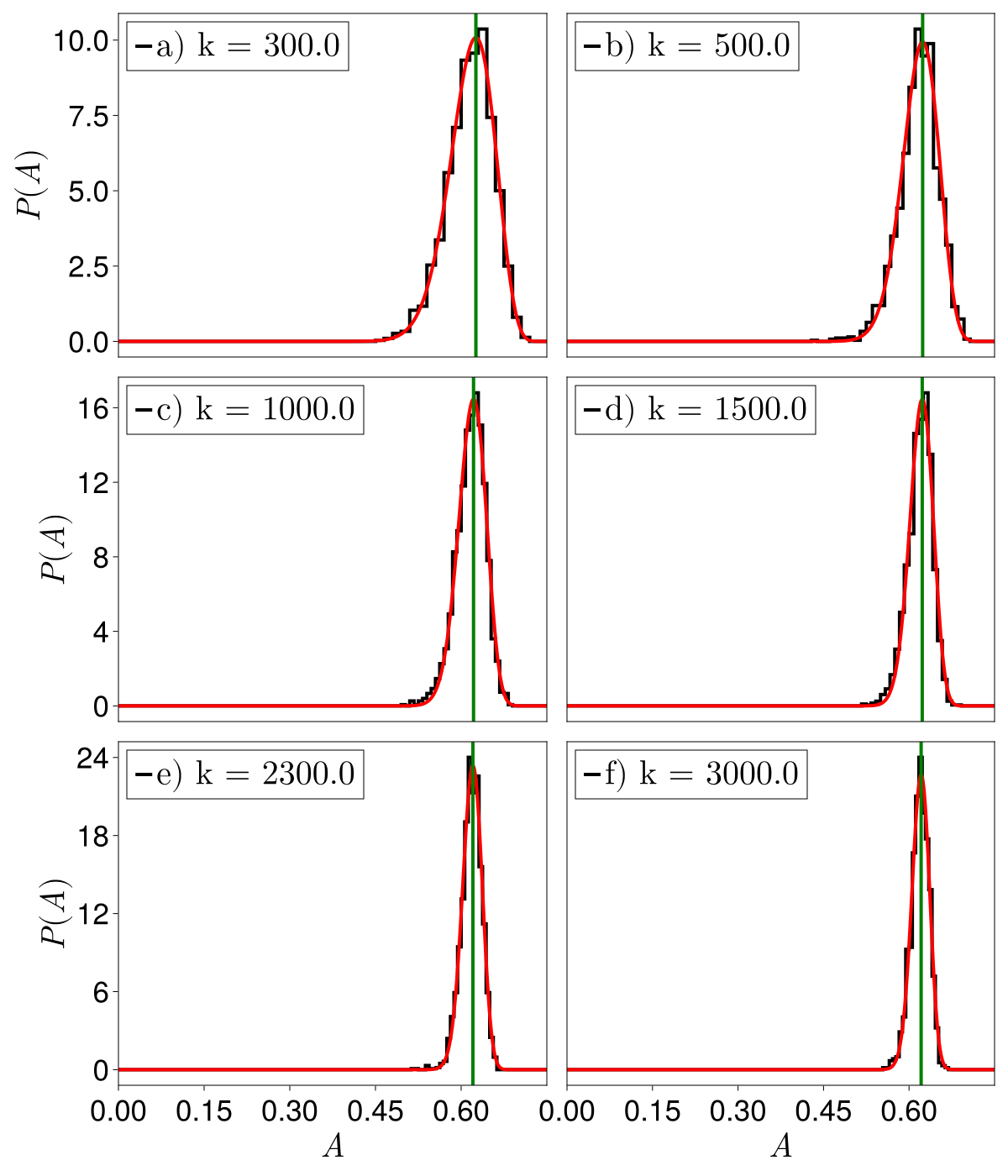}
    \caption{Distribution $P(A)$ of the localization measure $A$ derived from PH functions for the $C_3$ billiard at deformation parameter $a=0.2$ in the $m=0$ (GOE) symmetry sector. Panels (a)--(f) correspond to increasing wavenumbers $k=300,\,500,\,1000,\,1500,\,2300,$ and $3000$, respectively. Black histograms show numerical data obtained from PH functions, while solid red curves represent fits by the Beta distribution \eqref{beta_distr}. The vertical green line marks the maximum value of $P(A)$. With increasing $k$, the distribution becomes progressively narrower and more sharply peaked, indicating increasing phase--space delocalization and convergence toward the semiclassical ergodic limit.}
    \label{fig:PA_GOE}
\end{figure}

\subsection{Width of the Beta distribution vs. $k$}
\label{std_beta}

\begin{figure}
    \centering
    \includegraphics[width=\linewidth]{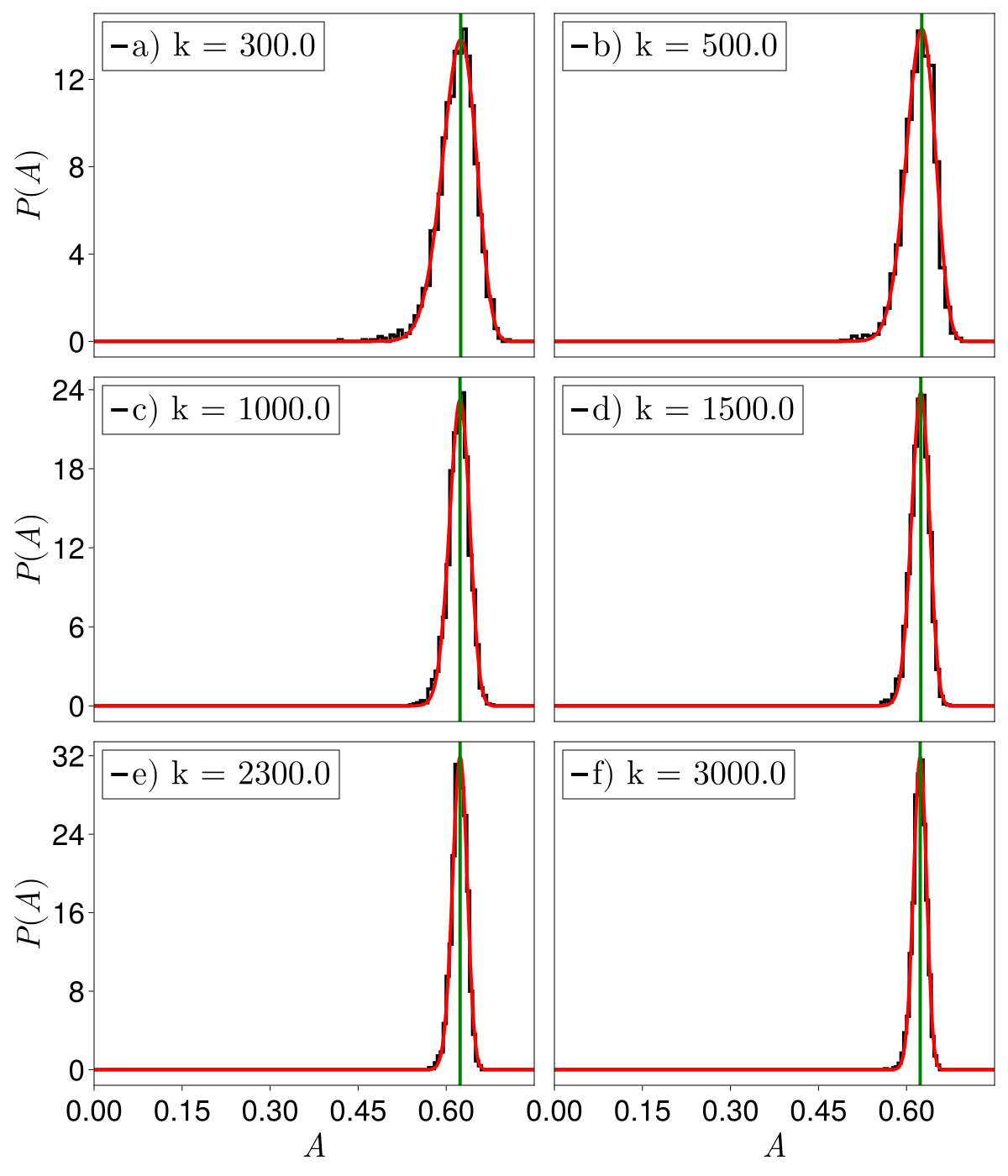}
    \caption{Same as Fig.~\ref{fig:PA_GOE}, but for the $m=1$ (GUE) symmetry sector of the $C_3$ billiard at $a=0.2$. The absence of time--reversal symmetry leads to systematically narrower distributions $P(A)$, consistent with enhanced phase--space delocalization compared to the GOE case.}
    \label{fig:PA_GUE}
\end{figure}

\begin{figure}
\centering

\includegraphics[width=\linewidth]{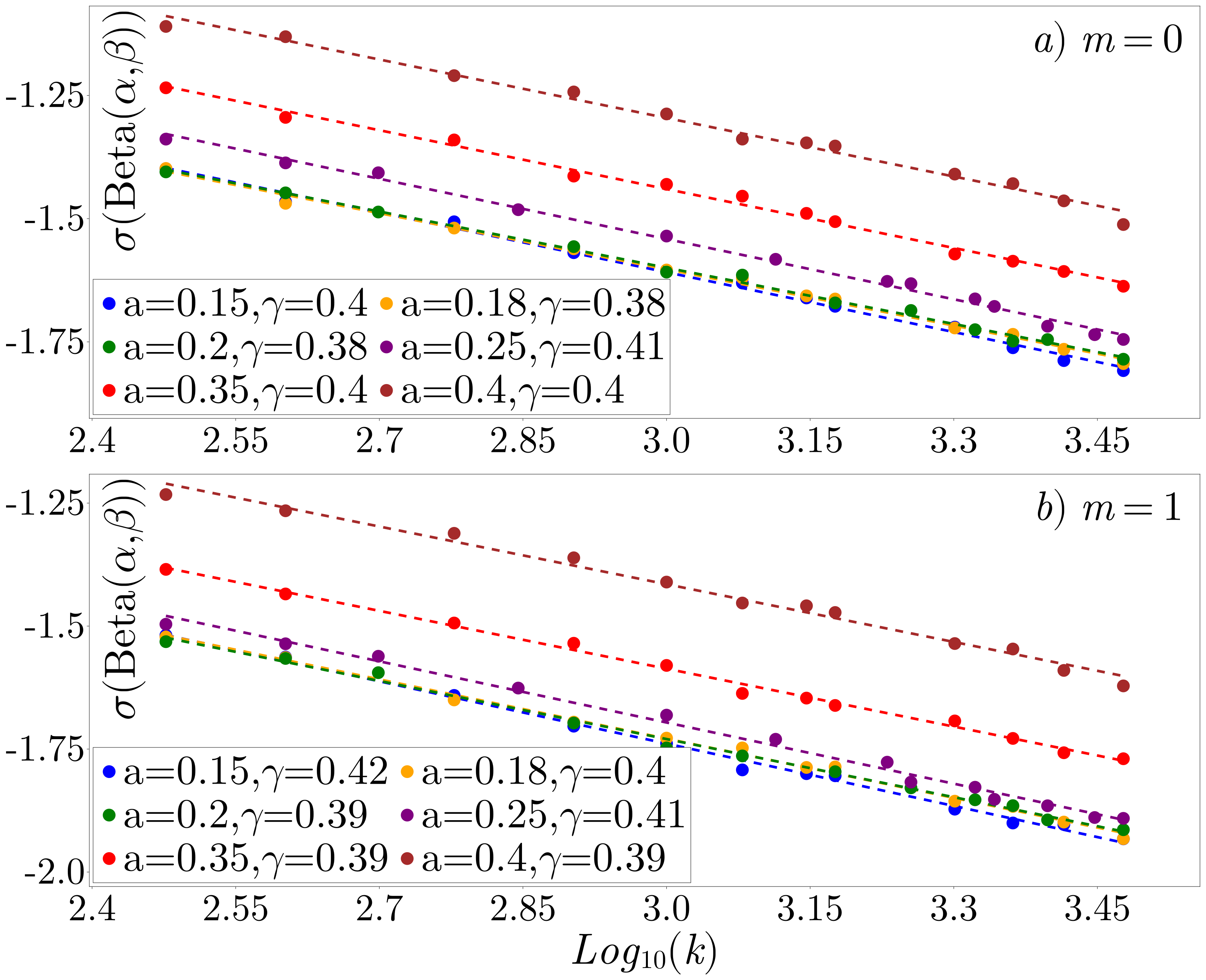}

\caption{Decay of the width $\sigma_\beta=\sigma(\mathrm{Beta}(\alpha,\beta))$
of the fitted Beta distribution describing the chaotic eigenstate statistics of
the localization measure $A$ for the deformations parameters $a \in [0.15,0.18,0.2,0.25,0.35,0.4]$. Shown is $\log_{10}\sigma_\beta$ versus
$\log_{10}k$ (decadic logarithms). (a) $m=0$ (GOE), (b) $m=1$ (GUE).
For all studied deformations $a = 0.15, 0.18, 0.20, 0.25, 0.35,$ and $0.40$, 
the data follow an approximately linear trend on the log--log scale, consistent with 
$\sigma_\beta \propto k^{-\gamma}$, indicating a progressive suppression of 
localization fluctuations in the semiclassical limit. 
The power--law behavior holds throughout the predominantly ergodic regime 
($0.05 \lesssim a \lesssim 0.31$), corresponding to $a = 0.15, 0.18, 0.20, 0.25$
and persisting in the mixed regime at $a = 0.35$ and $0.40$, where chaotic states exhibit only weak localization. Each dot represents the standard deviation of the Beta distribution fitted on an ensemble of $2000$ PH functions.}
\label{fig:std_beta_both}
\end{figure}

To quantify the approach to the semiclassical limit, we analyze the statistical
fluctuations of the localization--measure distribution obtained from the
PH representations of the eigenstates. For each chaotic eigenfunction,
the localization properties are characterized by fitting the $P(A)$
distribution with a Beta distribution.To this end, we examine the standard deviation ($\sigma$) which provides a direct measure of the width of the localization--measure distribution. The standard deviation of the Beta function is given by $\sigma(Beta(\alpha,\beta)) = \sqrt{Var(P(A)}$ where 

\begin{equation} \label{eq:std_beta}
    Var(P(A)) = A_0^2\frac{\alpha \beta}{(\alpha + \beta)^2(\alpha + \beta + 1)},
\end{equation}

\noindent is the variance of the fitted Beta distribution. For classically ergodic systems, Schnirelman's theorem \cite{shnirelman1974} (quantum ergodicity)
states that there exists a subsequence of eigenfunctions of density one whose 
semiclassical measures converge to the Liouville measure on the energy shell.
As a consequence, in billiards projections of these measures into configuration space yield
spatial equidistribution of the corresponding eigenfunctions. While the theorem
itself does not provide quantitative rates of convergence, semiclassical arguments
and random--wave models suggest that fluctuations of localization measures should
systematically decrease with increasing energy. This is a special, limiting case of the principle of semiclassical condensation (PUSC) of Wigner and Husimi functions \cite{Rob1998}.

Across all studied billiard geometries, we observe a clear algebraic decay of the
$\sigma(Beta(\alpha,\beta)) \propto k^{-\gamma}$ with exponents clustered around $\gamma =0.4$. This behavior is robust across symmetry sectors ($m=0,1$) and deformation parameters $a$. The result indicates a systematic self--averaging trend consistent with semiclassical expectations associated with quantum ergodicity \cite{husimi_zeros_and_spreading}. In physical terms, increasing energy progressively suppresses phase--space inhomogeneities. Importantly, the observed decay is algebraic rather than exponential, indicating that semiclassical convergence is governed by power--law scaling, as seen in Fig.\ref{fig:std_beta_both}.

\subsection{Approach to Quantum Ergodicity}
\label{conclusion_std_beta}

Using Poincaré--Husimi functions, we analyzed entropy-based localization
measures whose distributions are well described by Beta
distributions over a broad range of energies and deformation
parameters $a$. It is observed that the $\sigma (Beta(\alpha,\beta))$ for the $m=1$ irrep is consistently smaller than for $m=0$ at a given $k$. A characteristic feature observed across the studied cases is a power--law decay of the distribution width with
increasing wavenumber,

\begin{equation}
\sigma(\mathrm{Beta}(\alpha,\beta)) \propto k^{-\gamma}, \quad \gamma \approx 0.4.
\end{equation}

\noindent This behavior indicates a gradual reduction of localization
fluctuations toward the semiclassical limit. The extracted decay
exponents remain comparable across symmetry sectors and deformation
strengths within the predominantly ergodic regime. These findings are consistent with the expectations of quantum ergodicity, which predicts convergence of phase--space measures for
a subsequence of eigenstates of density one. Although the theorem
does not provide quantitative convergence rates, the observed
scaling suggests that localization fluctuations decrease
systematically with energy.

\begin{figure}
    \centering
    \includegraphics[width=\linewidth]{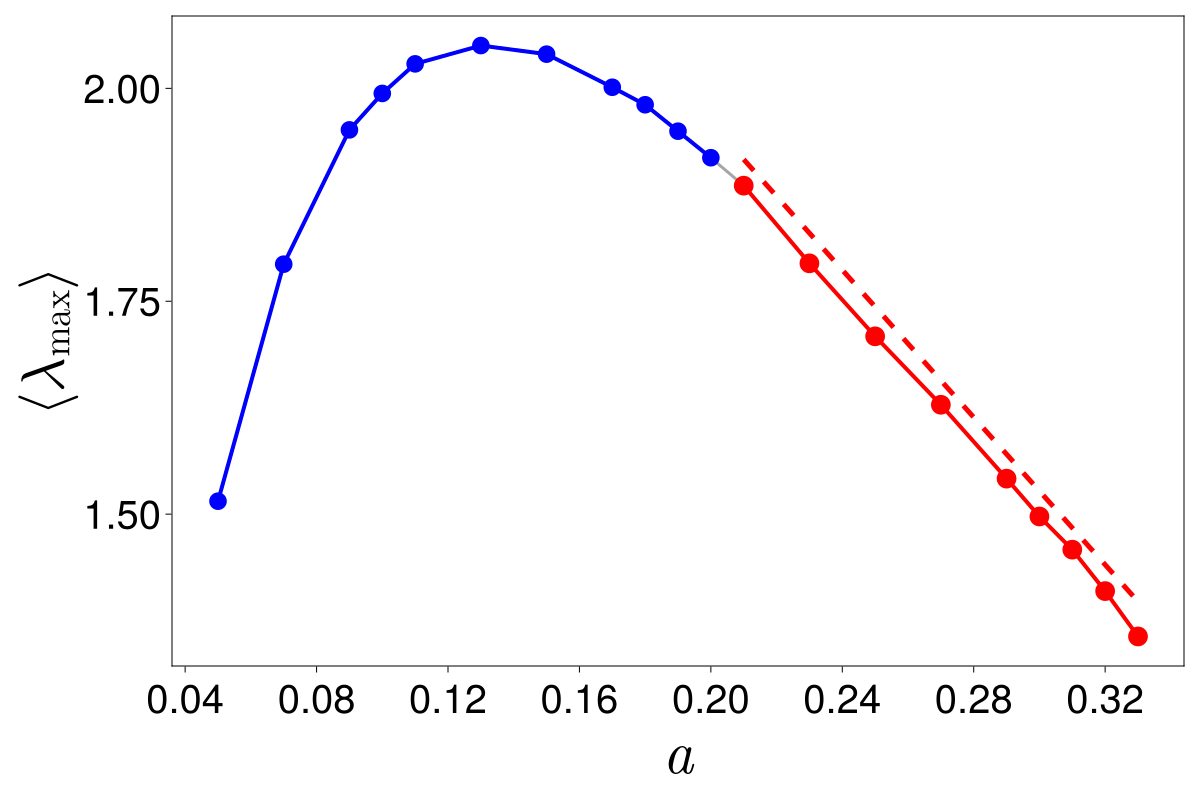}
    \caption{Average maximal Lyapunov exponent $\langle \lambda_{\max} \rangle$ as a function of the deformation parameter $a$. In the interval $a \in [0.2,0.34]$, the decrease of $\langle \lambda_{\max} \rangle$ is approximately linear. Comparing the $\langle \lambda_{\max} \rangle$ values for the given $a$ values with their accompanying intercept values $C$ from Fig.\ref{fig:std_beta_both} it is seen that an increase of $C$ is associated with a decrease of $\langle \lambda_{\max} \rangle$. For each initial condition in phase space, the trajectory was iterated until the finite-time maximal Lyapunov exponent stabilized within the prescribed numerical tolerance.}
    \label{fig:avg_lyap_vs_a}
\end{figure}

An interesting question that arises is whether a systematic relation exists between 
the average maximal Lyapunov exponent $\langle \lambda_{\max} \rangle$ 
within the chaotic region of phase space and the intercept $C$ obtained 
from the linear fits in the log--log representation

\[
\log_{10} \sigma(\mathrm{Beta}(\alpha,\beta)) 
= C - \gamma \log_{10} k.
\]

\noindent As seen in Fig.~\ref{fig:std_beta_both}, the intercept $C$ increases 
with the deformation parameter $a$, while simultaneously $\gamma \approx 0.4$ remains unchanged as $\langle \lambda_{\max} \rangle$ decreases. The interval 
$a \in [0.2,0.34]$ is particularly noteworthy, as  $\langle \lambda_{\max} \rangle$ exhibits an approximately linear  dependence on $a$ (Fig.\ref{fig:avg_lyap_vs_a}). This observation may provide further quantitative insight into the connection between classical instability and quantum localization fluctuations. Overall, the results show that fluctuation properties of localization measures provide information complementary to
conventional spectral diagnostics and offer a useful perspective on the trend toward semiclassical behavior.

\section{General Conclusions}
\label{general_conclusions}

We have presented a combined analysis of spectral correlations and eigenstate localization in the $C_3$--symmetric billiard introduced in Ref.~\cite{c3_init}. Using a contour--integral based approach and an explicit projection onto irreducible symmetry sectors, we obtained large spectra in the $m=0$ and $m=1$ subspaces and computed Poincar\'e--Husimi representations for large ensembles of eigenstates. 

On the spectral side we analyzed the nearest--neighbor spacing statistics showing excellent agreement with random--matrix universality when analyzed within each symmetry
sector with eigenvalues in the interval $k \in [5,3600]$: the real $m=0$ irreducible representation follows GOE statistics, while the complex $m=1$ irreducible representation follows GUE statistics, despite the time--reversal invariance of the underlying classical dynamics. Long--range correlations, quantified by the spectral rigidity $\Delta_3(L)$, agree with the corresponding GOE/GUE predictions up to the expected saturation scale controlled by the shortest relevant periodic orbit, consistent with semiclassical arguments.
Our main focus has been on phase--space eigenfunction localization in the chaotic regime. We characterized the phase--space structure of eigenstates using  Poincar\'e--Husimi functions and an
entropy localization measure $A$, constructed from said functions. For predominantly chaotic ensembles, the empirical distribution $P(A)$ is well described by a Beta distribution across a broad range of energies and geometry deformations. To quantify the approach toward the semiclassical regime, we tracked the width of the fitted distribution via $\sigma(\mathrm{Beta}(\alpha,\beta))$ and found a robust power--law decay $ \sigma(\mathrm{Beta}(\alpha,\beta)) \propto k^{-\gamma}$ with $\gamma \approx 0.4$ for all investigated deformations within the nearly fully ergodic interval. This trend indicates a systematic and quantifiable suppression of localization fluctuations with increasing energy, consistent with the qualitative expectations associated with quantum ergodicity \cite{shnirelman1974}. Beyn's method shows promise of being an extremely efficient method for calculating high--energy spectra of non--convex billiards with automatic verification of eigenvalue accuracy. To the best of our knowledge, it has not previously been applied to quantum billiard computations at this scale.

\section*{Acknowledgements}

This work was supported by the Slovenian Research and Innovation Agency (ARIS) under grants J1-4387 and P1-0306, and made use of the HPC VEGA supercomputer system under project S24O02-01.

\bibliography{main_final}

@PREAMBLE{
 "\providecommand{\noopsort}[1]{}" 
 # "\providecommand{\singleletter}[1]{#1}%" 
}

@BOOK{Stoe,
  AUTHOR =       {H.-J. St\"ockmann},
  TITLE =        {Quantum Chaos - An Introduction},
  PUBLISHER =    {Cambridge: Cambridge University Press},
  YEAR =         {1999},
}

@BOOK{Haake,
  AUTHOR =       {F. Haake},
  TITLE =        {Quantum Signatures of Chaos},
  PUBLISHER =    {Berlin: Springer},
  YEAR =         {2010},
}

@ARTICLE{Bunimovich2001,
   author       = "L. A. Bunimovich",
   year         = "2001",
   journal      = "Chaos",
   volume       = "11",
   pages        = "802-808",
}

@ARTICLE{LLR2021,
   author       = "\v{C}. Lozej and  D. Lukman and M. Robnik",
   year         = "2021",
   journal      = " Phys. Rev. E",
   volume       = "103",
   pages        = "012204",
}

@ARTICLE{BLR2018,
   author       = "B. Batisti\'c and \v{C}. Lozej and M. Robnik",
   year         = "2018",
   journal      = "Nonlinear Phenomena in Complex Systems (Minsk)",
   volume       = "21",
   pages        = "225",
}

@ARTICLE{BLR2020,
   author       = "B. Batisti\'c and \v{C}. Lozej and M. Robnik",
   year         = "2020",
   journal      = "Nonlinear Phenomena in Complex Systems (Minsk)",
   volume       = "23",
   pages        = "17",
}

@ARTICLE{BLR2019B,
   author       = "B. Batisti\'c and \v{C}. Lozej and M. Robnik",
   year         = "2019",
   journal      = "Phys. Rev. E",
   volume       = "100",
   pages        = "062208",
}

@ARTICLE{Rob1998,
   author       = "M. Robnik",
   year         = "1998",
   journal      = "Nonlinear Phenomena in Complex Systems (Minsk)",
   volume       = "1",
   pages        = "1",
}

@BOOK{Mehta,
  AUTHOR =       {M.~L. Mehta},
  TITLE =        {Random Matrices},
  PUBLISHER =    {Boston: Academic Press},
  YEAR =         {1991},
}

@ARTICLE{BGS1984,
   author       = "O. Bohigas and M.~J. Giannoni and C. Schmit",
   journal      = "Phys. Rev. Lett.",
   volume       = "52", 
   pages        = "1",
   year         = "1984",
}

@ARTICLE{Cas1980,
   author       = "G. Casati and F. Valz-Gris and I. Guarnieri",
   journal      = "Lett. Nuovo Cimento",
   volume       = "28", 
   pages        = "279",
   year         = "1980",
}

@ARTICLE{Berry1985,
   author       = "M.~V. Berry",
   journal      = "Proc. Roy. Soc. Lond. A",
   volume       = "400", 
   pages        = "229",
   year         = "1985",
}

@ARTICLE{BerTab1977,
   author       = "M.~V. Berry and M. Tabor",
   journal      = "Proc. Roy. Soc. Lond. A",
   volume       = "356", 
   pages        = "375",
   year         = "1977",
}

@article{BatManRob2013,
  author={B. Batisti{\'c} and T. Manos and M. Robnik},
  title={The intermediate level statistics in dynamically localized chaotic eigenstates},
  journal={EPL},
  volume={102},
  number={5},
  pages={50008},
  year={2013},
}

@ARTICLE{Wig1932,
   author       = {E. Wigner},
   journal      = "Phys. Rev.",
   volume       = "40", 
   pages        = "749",
   year         = "1932",
}

@ARTICLE{Hus1940,
   author       = {K. Husimi},
   journal      = "Proc. Phys. Math. Soc. Jpn.",
   volume       = "22", 
   pages        = "264",
   year         = "1940",
}

@ARTICLE{Bro1973,
   author       = "T. A. Brody ",
   journal      = "Lett. Nuovo Cimento",
   volume       = "7", 
   pages        = "482",
   year         = "1973",
}

@article{VebProRob2007,
  author={G Veble and T Prosen and M Robnik},
  title={Expanded boundary integral method and chaotic time-reversal doublets in quantum billiards},
  journal={New J. Phys.},
  volume={9},
  number={1},
  pages={15},
  year={2007},
}

@article{VerSar1995,
  title = {Calculation by scaling of highly excited states of billiards},
  author = {Vergini, E. and Saraceno, M.},
  journal = {Phys. Rev. E},
  volume = {52},
  issue = {3},
  pages = {2204--2207},
  year = {1995},
}

@ARTICLE{Izr1989,
   author       = "F. M. Izrailev",
   journal      = "J. Phys. A: Math. Gen.",
   volume       = "22", 
   pages        = "865",
   year         = "1989",
}

@ARTICLE{Rob1983,
   author       = "M. Robnik",
   journal      = "J. Phys. A: Math. Gen.",
   volume       = "16", 
   pages        = "3971",
   year         = "1983",
}

@ARTICLE{Heller1984,
    author      = "E. J. Heller",
    journal     = "Phys. Rev. Lett.",
    volume      = "53",
    pages       = "1515",
    year        = "1984",
}

@BOOK{Steiner1994,
  AUTHOR =       {F. Steiner},
  TITLE =        {in Universit\"at Hamburg 1994: Schlaglichter der Forschung zum 75. Jahrestag, ed. by R. Ansorge, p. 543},
  PUBLISHER =    {Hamburg: Reimer},
  YEAR =         {1994},
}

@ARTICLE{Richter1998,
    author      = "H. Alt and A. Backer and C. Dembowski and H.-D. Graf and R. Hofferbert and H. Rehfeld and A. Richter",
    journal     = "Phys. Rev. E",
    volume      = "58",
    pages       = "1737",
    year        = "1998",
}

@article{berry1977,
  title={Regular and irregular semiclassical wavefunctions},
  author={Berry, Michael V},
  journal={Journal of Physics A: Mathematical and General},
  volume={10},
  number={12},
  pages={2083},
  year={1977},
  publisher={IOP Publishing}
}

@article{shnirelman1974,
  title={Uspekhi Matem},
  author={Shnirelman, B},
  journal={Nauk},
  volume={29},
  pages={181},
  year={1974}
}

@incollection{voros1979,
  title={Semi-classical ergodicity of quantum eigenstates in the Wigner representation},
  author={Voros, Andr{\'e}},
  booktitle={Stochastic behavior in classical and quantum Hamiltonian systems},
  pages={326--333},
  year={1979},
  publisher={Springer}
}

@article{lozej2022triangles,
  title={Quantum chaos in triangular billiards},
  author={Lozej, {\v{C}}rt and Casati, Giulio and Prosen, Toma{\v{z}}},
  journal={Physical Review Research},
  volume={4},
  number={1},
  pages={013138},
  year={2022},
  publisher={APS}
}

@article{kress,
    author={Kress, Rainer},
    title={Boundary Integral Equations in time-harmonic acoustic scattering},
    journal={Mathematical and Computer Modelling},
    volume={15},
    number={3-5},
    pages={229-243},
    year={1991},
    publisher={Pergamon Press plc}
}

@article{Sinai1970,
    author={Ya. G. Sinai},
    title={Dynamical systems with elastic reflections. Ergodic properties of dispersing billiards},
    journal={Russian Mathematical Surveys},
    year={1970},
    volume={25},
    pages={137-189},
    number={2},
}

@article{Backer2003,
  title={Poincar\'e Husimi representation of eigenstates in quantum billiards},
  author={B\"acker, A. and F\"urstberger, S. and Schubert, R.},
  journal={Phys. Rev. E},
  volume={70},
  issue={3},
  pages={036204},
  numpages={10},
  year={2004},
  publisher={American Physical Society},
}

@article{Yan_FPUT_mixed,
  title = {Chaos and quantization of the three-particle generic Fermi-Pasta-Ulam-Tsingou model. II. Phenomenology of quantum eigenstates},
  author = {Yan, Hua and Robnik, Marko},
  journal = {Phys. Rev. E},
  volume = {109},
  issue = {5},
  pages = {054211},
  year = {2024},
  publisher = {American Physical Society},
}

@article{Veble_Robnik_integrable,
year = {1998},
publisher = {},
volume = {31},
number = {20},
pages = {4669},
author = {Marko Robnik and Gregor Veble},
title = {On spectral statistics of classically integrable systems},
journal = {Journal of Physics A: Mathematical and General},
}

@article{Jiang_laser,
  author  = {Jiang, Xue-Feng and Zou, Chang-Ling and Wang, Lei and Gong, Qi-Huang and Xiao, Yun-Feng},
  title   = {Whispering-gallery microcavities with unidirectional laser emission},
  journal = {Laser \& Photonics Reviews},
  volume  = {10},
  pages   = {40--61},
  year    = {2016},
}

@article{experiment_microwave_tunneling,
  title = {Experimental investigations of chaos-assisted tunneling in a microwave annular billiard},
  author = {Hofferbert, R. and Alt, H. and Dembowski, C. and Gr\"af, H.-D. and Harney, H. L. and Heine, A. and Rehfeld, H. and Richter, A.},
  journal = {Phys. Rev. E},
  volume = {71},
  pages = {046201},
  year = {2005},
  publisher = {American Physical Society},
}

@article{beyn_contour,
    title = {An integral method for solving nonlinear eigenvalue problems},
    journal = {Linear Algebra and its Applications},
    volume = {436},
    number = {10},
    pages = {3839-3863},
    year = {2012},
    author = {Wolf-Jürgen Beyn},
}

@article{sakurai_contour,
    author  = {Akira Imakura and Lei Du and Tetsuya Sakurai},
    title   = {A map of contour integral-based eigensolvers for solving generalized eigenvalue problems},
    journal = {Japan Journal of Industrial and Applied Mathematics},
    year    = {2016},
    volume  = {33},
    number  = {3},
    pages   = {721--750},
}

@book{keldysh_theorem,
  author    = {R. Mennicken and M. M{\"o}ller},
  title     = {Non-Self-Adjoint Boundary Eigenvalue Problems},
  series    = {North-Holland Mathematics Studies},
  volume    = {192},
  publisher = {Elsevier},
  year      = {2003},
  isbn      = {9780080537733}
}

@article{backer_boundary,
    year = {2002},
    month = {nov},
    volume = {35},
    number = {48},
    pages = {10293},
    author = {A Bäcker and S Fürstberger and R Schubert and F Steiner},
    title = {Behaviour of boundary functions for quantum billiards},
    journal = {Journal of Physics A: Mathematical and General},
}

@article{lyapunov_exponent,
    author = {A. M. Lyapunov},
    title = {The general problem of the stability of motion},
    journal = {International Journal of Control},
    volume = {55},
    number = {3},
    pages = {531--534},
    year = {1992},
    publisher = {Taylor \& Francis},
}

@article{dietz_c3,
  title = {Test of a numerical approach to the quantization of billiards},
  author = {Dietz, B. and Heine, A. and Heuveline, V. and Richter, A.},
  journal = {Phys. Rev. E},
  volume = {71},
  issue = {2},
  pages = {026703},
  numpages = {6},
  year = {2005},
  month = {Feb},
  publisher = {American Physical Society},
}

@article{c3_init,
  title = {Phase Shift Experiments Identifying Kramers Doublets in a Chaotic Superconducting Microwave Billiard of Threefold Symmetry},
  author = {Dembowski, C. and Dietz, B. and Gr\"af, H.-D. and Heine, A. and Leyvraz, F. and Miski-Oglu, M. and Richter, A. and Seligman, T. H.},
  journal = {Phys. Rev. Lett.},
  volume = {90},
  issue = {1},
  pages = {014102},
  numpages = {4},
  year = {2003},
  month = {Jan},
  publisher = {American Physical Society},
}

@article{benettin_normalization_1,
  author  = {Benettin, Giancarlo and Galgani, Luigi and Giorgilli, Antonio and Strelcyn, Jean-Marie},
  title   = {Lyapunov Characteristic Exponents for Smooth Dynamical Systems and for Hamiltonian Systems. Part I: Theory},
  journal = {Meccanica},
  year    = {1980},
  volume  = {15},
  number  = {1},
  pages   = {9--20},
}

@article{benettin_normalization_2,
  author  = {Benettin, Giancarlo and Galgani, Luigi and Giorgilli, Antonio and Strelcyn, Jean-Marie},
  title   = {Lyapunov Characteristic Exponents for Smooth Dynamical Systems and for Hamiltonian Systems; A Method for Computing All of Them. Part 2: Numerical Application},
  journal = {Meccanica},
  year    = {1980},
  volume  = {15},
  number  = {1},
  pages   = {21--30},
}

@article{lyapunov_billiards,
  title = {Lyapunov instability in a system of hard disks in equilibrium and nonequilibrium steady states},
  author = {Dellago, Ch. and Posch, H. A. and Hoover, W. G.},
  journal = {Phys. Rev. E},
  volume = {53},
  issue = {2},
  pages = {1485--1501},
  numpages = {0},
  year = {1996},
  month = {Feb},
  publisher = {American Physical Society},
}

@InProceedings{backer_BIM,
    author="B{\"a}cker, Arnd",
    editor="Esposti, Mirko Degli
    and Graffi, Sandro",
    title="Numerical Aspects of Eigenvalue and Eigenfunction Computations for Chaotic Quantum Systems",
    booktitle="The Mathematical Aspects of Quantum Maps",
    year="2003",
    publisher="Springer Berlin Heidelberg",
    address="Berlin, Heidelberg",
    pages="91--144",
}

@book{Irreps_ref,
  author    = {Michael Tinkham},
  title     = {Group Theory and Quantum Mechanics},
  publisher = {McGraw-Hill},
  year      = {1964},
}

@article{c4,
  title = {Quantization and interference of a quantum billiard with fourfold rotational symmetry},
  author = {Li, Zi-Yuan and Huang, Liang},
  journal = {Phys. Rev. E},
  volume = {101},
  issue = {6},
  pages = {062201},
  numpages = {11},
  year = {2020},
  month = {Jun},
  publisher = {American Physical Society},
}

@article{dietz_threefold,
  author  = {Dietz, Barbara},
  title   = {Relativistic Quantum Billiards with Threefold Rotational Symmetry: Exact, Symmetry-Projected Solutions for Equilateral Neutrino Billiard},
  journal = {Acta Physica Polonica A},
  year    = {2021},
  volume  = {140},
  number  = {6},
  pages   = {473--486},
}

@article{dietz_poly_basis_method_c3,
  title = {Test of a numerical approach to the quantization of billiards},
  author = {Dietz, B. and Heine, A. and Heuveline, V. and Richter, A.},
  journal = {Phys. Rev. E},
  volume = {71},
  issue = {2},
  pages = {026703},
  numpages = {6},
  year = {2005},
  month = {Feb},
  publisher = {American Physical Society},
}

@article{husimi_zeros_and_spreading,
year = {1990},
month = {may},
publisher = {},
volume = {23},
number = {10},
pages = {1765},
author = {P Leboeuf and A Voros},
title = {Chaos-revealing multiplicative representation of quantum eigenstates},
journal = {Journal of Physics A: Mathematical and General},
}

@article{dietz_c3_new,
  title = {Microwave photonic crystals, graphene, and honeycomb-kagome billiards with threefold symmetry: Comparison with nonrelativistic and relativistic quantum billiards},
  author = {Zhang, Weihua and Dietz, Barbara},
  journal = {Phys. Rev. B},
  volume = {104},
  issue = {6},
  pages = {064310},
  numpages = {25},
  year = {2021},
  month = {Aug},
  publisher = {American Physical Society},
}

@article{prosen_shortest_PO,
year = {1993},
publisher = {},
volume = {26},
number = {10},
pages = {2371},
author = {T Prosen and M Robnik},
title = {Energy level statistics in the transition region between integrability and chaos},
journal = {Journal of Physics A: Mathematical and General},
}

@book{abramowitz,
    author = {Abramowitz, M. and Stegun, I.A.},
    title = {Handbook of Mathematical Functions with Formulas, Graphs, and Mathematical Tables},
    publisher = {Dover Publications},
    year = {1964}
}

@misc{hank106,
  author       = {Greengard, L.},
  title        = {mpspack: hank106.f},
  year         = {2026},
  howpublished = {\href{https://github.com/ahbarnett/mpspack/blob/master/@utils/hank106.f}{GitHub source file}},
  note         = {Accessed: 2026-02-16}
}

@article{robnik1986_GUE,
    year = {1986},
    publisher = {IOPscience},
    volume = {19},
    number = {5},
    pages = {669},
    author = {M Robnik and M V Berry},
    title = {False time-reversal violation and energy level statistics: the role of anti-unitary symmetry},
    journal = {Journal of Physics A: Mathematical and General},
}

@article{stockmann_microwave_billiards_2022,
    year = {2022},
    publisher = {IOP Publishing},
    volume = {55},
    number = {38},
    pages = {383001},
    author = {Stöckmann, Hans-Jürgen and Kuhl, Ulrich},
    title = {Microwave studies of the spectral statistics in chaotic systems},
    journal = {Journal of Physics A: Mathematical and Theoretical},
}

@InProceedings{robnik_proceedigs_1986_antiunitary,
    author="Robnik, M.",
    title="Antiunitary symmetries and energy level statistics",
    booktitle="Quantum Chaos and Statistical Nuclear Physics",
    year="1986",
    publisher="Springer Berlin Heidelberg",
    address="Berlin",
    pages="120--130",
}

@article{robnik_review_2023,
  author  = {Robnik, M.},
  title   = {Recent Developments in Quantum Chaos of Mixed-Type Systems: A Mini Review},
  journal = {Nonlinear Phenomena in Complex Systems},
  volume  = {26},
  number  = {3},
  pages   = {209--224},
  year    = {2023}
}

@article{batistic_robnik_localization_eigenstates,
  title = {Quantum localization of chaotic eigenstates and the level spacing distribution},
  author = {Batisti\ifmmode \acute{c}\else \'{c}\fi{}, Benjamin and Robnik, Marko},
  journal = {Phys. Rev. E},
  volume = {88},
  issue = {5},
  pages = {052913},
  numpages = {7},
  year = {2013},
  publisher = {American Physical Society},
}

@article{husimi_original,
  title = {Quantum Poincar\'e sections for two-dimensional billiards},
  author = {Crespi, Bruno and Perez, Gabriel and Chang, Shau-Jin},
  journal = {Phys. Rev. E},
  volume = {47},
  issue = {2},
  pages = {986--991},
  numpages = {0},
  year = {1993},
  publisher = {American Physical Society},
}

\appendix

\newpage

\section{Numerical details: Beyn contour method}
\label{app:beyn}

\begin{figure}[t]
    \centering
    \includegraphics[width=1.0\linewidth]{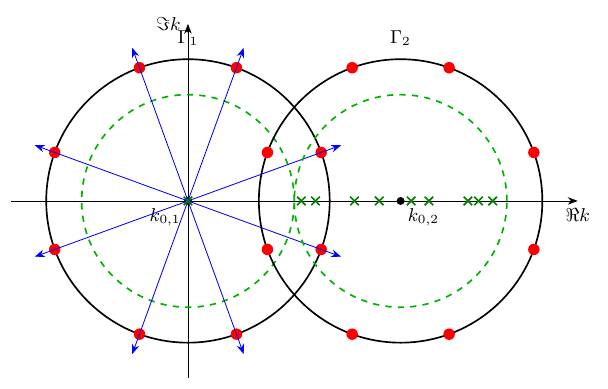}
    \caption{Two successive complex contour centers $k_{0,1}$ and $k_{0,2}$ with their associated circular integration contours (black). Since the accuracy of the projected eigenvalues obtained from diagonalizing \ref{eq:B_similar_clean2} deteriorates near the contour boundary,adjacent contours are chosen to overlap slightly. Only eigenvalues lying within the inner, non-overlapping regions (green dashed circles) are retained, thereby preserving numerical accuracy. The contour radius is estimated using Weyl's law, supplemented by a small safety margin to ensure that no eigenvalues are missed. In practice, we employ a trapezoidal discretization with $N=50$ quadrature nodes and parameterize each contour as $\Gamma_i = \{ k_{0,i} + R \exp{(j \phi)} |\phi=\frac{2\pi}{N},j=1,...,N-1\}$ where $R$ is chosen such that at most $150$ eigenvalues are expected within the contour In the computation the contour radius was restricted to $R \le 0.5$ (high $k$ range), since larger values led to spurious solutions (manifested as unresolved small singular values, where the numerical threshold for zero becomes ambiguous) and reduced precision (see Integral algorithm $1.$, step $4$ in \cite{beyn_contour}). This discretization is sufficient for the moment matrix $A_0$ to become unambiguously numerically rank-deficient, indicating that all physical eigenvalues enclosed by the contour have been captured.}
    \label{fig:beyn_contour_sketch}
\end{figure}

To compute high--lying eigenvalues of the Helmholtz operator in billiard domains,
we employ a contour--integral method for nonlinear eigenvalue problems based on
the framework of Beyn \cite{beyn_contour,sakurai_contour}. The primary motivation for adopting Beyn’s contour–integral method is that within the deformation–parameter range of interest, the billiard boundary becomes non-convex and exhibits strongly varying curvature. Under these conditions, the traditional Vergini–Saraceno method \cite{VerSar1995} is no longer directly applicable. Beyn's method similarly to EBIM \cite{VebProRob2007} gives eigenvalues with a small imaginary component. This residual imaginary part provides a practical indicator of numerical accuracy, reflecting discretization errors associated with the finite-dimensional approximation of the Fredholm operator and highlighting spurious solutions.

Beyn's method can be used to find all solutions $(z,v(z))$ of nonlinear operators satisfying the following equation $T(z)v(z)=0$ where $z$ is typically a complex parameter and $v(z)$ the solution vector. Since in BIM this is exactly the equation we are solving \cite{backer_BIM}, Beyn's method is applicable. Let $T(k)$ denote the nonlinear Fredholm operator arising from the boundary integral formulation of the Helmholtz equation. For $k$ inside a contour $\Gamma$ in the complex plane, the inverse operator admits the following expansion due to Keldysh's theorem \cite{keldysh_theorem}

\begin{equation} \label{eq:keldysh}
T(z)^{-1}
=
\sum_{n=1}^{N_\Gamma} \frac{1}{z-\lambda_n}\, v_n w_n^\dagger
+ R(z),
\end{equation}

\noindent where $\{\lambda_n\}$ are $N_\Gamma$ eigenvalues enclosed by $\Gamma$,
$v_n$ and $w_n$ are the corresponding right and left eigenvectors of $T(\lambda_n)$, with the normalization $w_n^\dagger T'(\lambda_n)v_n = 1$ (see Eq.(6,7) in \cite{beyn_contour}),
and $R(z)$ is analytic inside $\Gamma$. This representation forms the theoretical basis of contour--integration
methods for nonlinear eigenvalue problems. The robustness of the method is guaranteed due to interior Helmholtz problems with closed boundaries having only real eigenvalues, implying that all the poles of \eqref{eq:keldysh} are on the real axis and therefore we are never evaluating nearly singular Fredholm matrices on the contour (although care is taken when Fredholm matrices are being evaluated near the intersection of the contour with the real axis since an eigenvalue could be lying close). For an analysis of how $N_\Gamma$ determines eigenvalue accuracy see Fig.\ref{fig:res_vs_N}. Following Beyn the following lowest contour moments are constructed:

\begin{figure}
    \centering
    \includegraphics[width=1.1\linewidth]{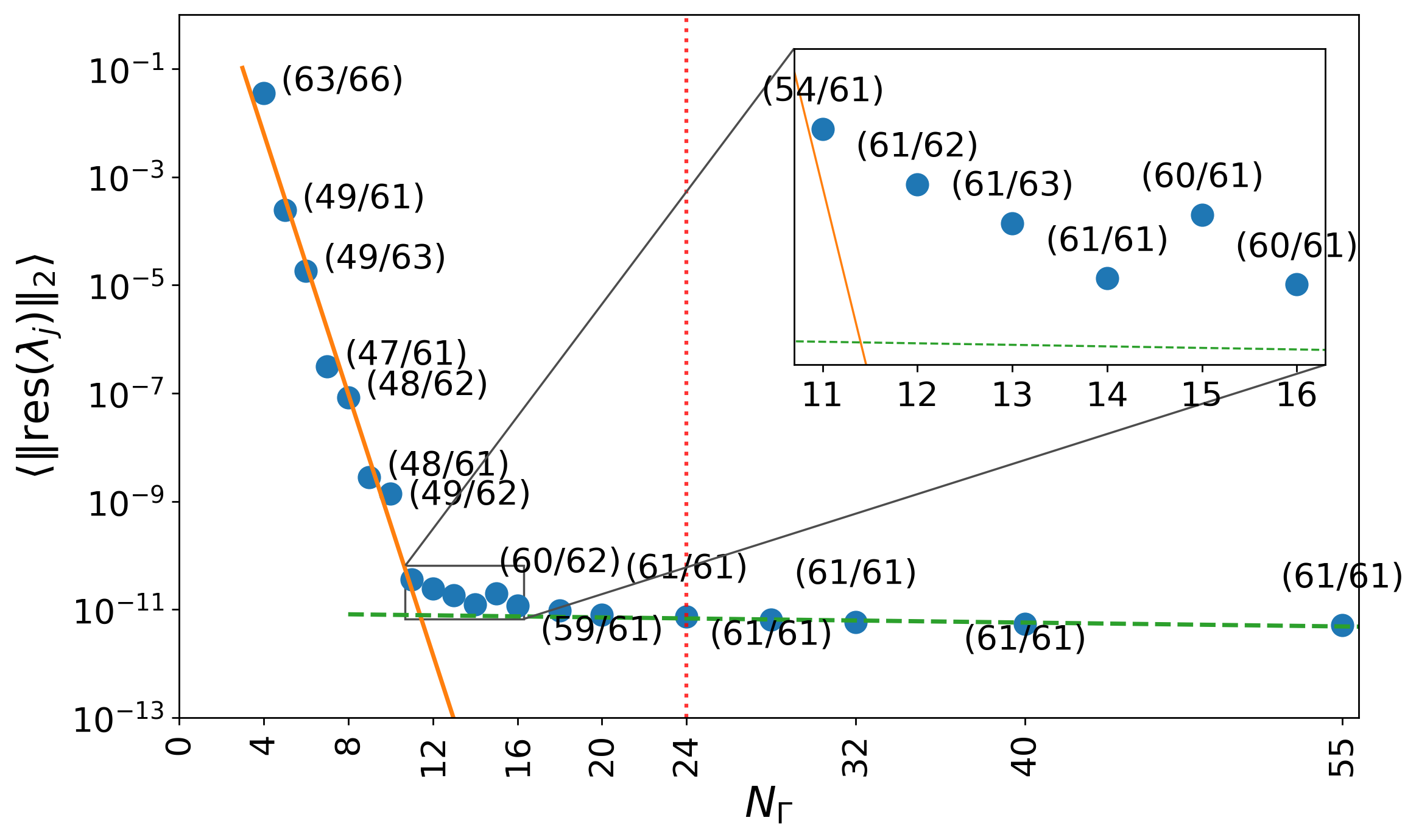}
    \caption{Log–lin plot of the average residual 
    $\langle \|\mathrm{res}(\lambda_j)\|_{2} \rangle
    = \langle \|T(\lambda_j)v(\lambda_j)\|_{2} \rangle$
    for $a=0.2$ and $m=0$ as a function of the contour discretization
    $N_{\Gamma}$. 
    The contour ($R=0.4$) encloses the interval
    $k\in(1404.0,1404.8)$ containing 61 non-spurious eigenvalues,
    over which the average is taken. 
    An exponential decrease of the residual is observed up to
    $N_{\Gamma}\approx12$, followed by saturation at
    $\langle \|\mathrm{res}(\lambda_j)\|_{2} \rangle
    \approx 8\times10^{-12}$.
    The labels above each point show 
    $(\mathrm{keep}/\mathrm{found})$, where
    ``found'' denotes the total number of eigenvalues detected inside the contour,
    while ``keep'' denotes the number of non-spurious eigenvalues that pass the
    post-processing filtering criteria and are retained for analysis.
    Although the residual magnitude appears stable for
    $N_{\Gamma}\in(12,24)$, it does not find all the eigenvalues, as $1$ or $2$ eigenvalues close to the contour edges are lost. Therefore only at $N_{\Gamma}=24$ (vertical dashed line) and beyond do we find all the eigenvalues inside the contour.
    For the full spectral range $k\in(5,3600)$ a conservative choice
    $N_{\Gamma}=50$ was adopted, yielding uniformly small residuals
    of order $10^{-10}$.}
    \label{fig:res_vs_N}
\end{figure}

\begin{equation}
A_0
=
\frac{1}{2\pi i}
\oint_\Gamma T(z)^{-1} V \, \mathrm{d}z,
\qquad
A_1
=
\frac{1}{2\pi i}
\oint_\Gamma z\, T(z)^{-1} V \, \mathrm{d}z,
\label{eq:contour_moments}
\end{equation}

\noindent where $V$ is a probing matrix with random columns. Inserting \eqref{eq:keldysh} into the definition of $A_0$ gives

\begin{equation}
\begin{aligned}
A_0
&=
\frac{1}{2\pi i}\oint_\Gamma T(z)^{-1}V\,dz \\
&=
\frac{1}{2\pi i}\oint_\Gamma
\left(
\sum_{n=1}^{N_\Gamma}\frac{1}{z-\lambda_n}\,v_n w_n^\dagger
+R(z)
\right)V\,dz \\
&=
\sum_{n=1}^{N_\Gamma}
\left[
\frac{1}{2\pi i}\oint_\Gamma \frac{1}{z-\lambda_n}\,dz
\right]
v_n\,(w_n^\dagger V)
\;+\;
\frac{1}{2\pi i}\oint_\Gamma R(z)V\,dz .
\end{aligned}
\label{eq:A0_expand}
\end{equation}

\noindent Since $R(z)V$ is analytic on and inside $\Gamma$, Cauchy's theorem gives
\(
\oint_\Gamma R(z)V\,dz=0.
\)
Moreover, by the residue theorem,
\(
\frac{1}{2\pi i}\oint_\Gamma \frac{1}{z-\lambda_n}\,dz=1
\)
for each $\lambda_n$ inside $\Gamma$. Therefore

\begin{equation}
A_0
=
\sum_{n=1}^{N_\Gamma} v_n\,(w_n^\dagger V).
\label{eq:A0_result}
\end{equation}

\noindent Similarly we have for the first moment

\begin{equation}
\begin{aligned}
A_1
&=
\frac{1}{2\pi i}\oint_\Gamma z\,T(z)^{-1}V\,dz \\
&=
\frac{1}{2\pi i}\oint_\Gamma
z\left(
\sum_{n=1}^{N_\Gamma}\frac{1}{z-\lambda_n}\,v_n w_n^\dagger
+R(z)
\right)V\,dz \\
&=
\sum_{n=1}^{N_\Gamma}
\left[
\frac{1}{2\pi i}\oint_\Gamma \frac{z}{z-\lambda_n}\,dz
\right]
v_n\,(w_n^\dagger V)
\;+\;
\frac{1}{2\pi i}\oint_\Gamma z\,R(z)V\,dz .
\end{aligned}
\label{eq:A1_expand}
\end{equation}

\noindent Again, $zR(z)V$ is analytic inside $\Gamma$, so its contour integral vanishes. For the pole terms we use the identity $\frac{z}{z-\lambda_n}=1+\frac{\lambda_n}{z-\lambda_n}$, hence $\frac{1}{2\pi i}\oint_\Gamma \frac{z}{z-\lambda_n}\,dz = \lambda_n$.

Therefore

\begin{equation}
A_1
=
\sum_{n=1}^{N_\Gamma} \lambda_n\, v_n\,(w_n^\dagger V).
\label{eq:A1_result}
\end{equation}

\noindent The common term gets relabeled as $\alpha^{\dagger}_n=w_n^{\dagger}V$

\begin{equation}
A_0=\sum_{n=1}^{r} v_n \alpha_n^\dagger,
\qquad
A_1=\sum_{n=1}^{r} \lambda_n\, v_n \alpha_n^\dagger,
\label{eq:moments_rank1}
\end{equation}

\noindent where $r=N_\Gamma$ is the number of eigenvalues inside $\Gamma$ (all unique since we are in a given irreducible representation). Introducing the matrices

\begin{equation}
\begin{aligned}
&\Phi = [v_1,\dots,v_r]\in\mathbb{C}^{N\times r}, \\
&\Lambda = \mathrm{diag}(\lambda_1,\dots,\lambda_r)\in\mathbb{C}^{r\times r},
\\
&C =
\begin{pmatrix}
\alpha_1^\dagger\\
\vdots\\
\alpha_r^\dagger
\end{pmatrix}
\in\mathbb{C}^{r\times L},
\end{aligned}
\end{equation}

\noindent where $r = \mathrm{rank}(A_0)$ by construction, $N$ is the dimension of $T(z)$, and $L$ is the number of expected eigenvalues in the contour with additional padding (due to the fact that there are fluctuations around the smooth part of Weyl's law we reasonably increase the probing size of the random matrix $V$ as to not lose potential eigenvalues.) Therefore \eqref{eq:moments_rank1} can be written compactly as

\begin{equation}
A_0 = \Phi C,
\qquad
A_1 = \Phi \Lambda C.
\label{eq:A0A1_fact_clean}
\end{equation}

\noindent We compute the singular value decomposition

\begin{equation}
A_0 = U \Sigma W^\dagger,
\qquad
U \in \mathbb{C}^{N\times r},\ 
\Sigma \in \mathbb{R}^{r\times r},\ 
W \in \mathbb{C}^{L\times r},
\label{eq:A0_svd_thin_clean}
\end{equation}

\noindent and since $A_0 = \Phi C$ has guaranteed rank $r$, the columns of $U$ form an orthonormal basis of the eigenspace spanned by the vectors $\{v_n\}$, implying that the matrix

\begin{equation}
S = U^\dagger \Phi \in \mathbb{C}^{r\times r},
\label{eq:S_def_clean2}
\end{equation}

\noindent is invertible. Constructing now the reduced matrix

\begin{equation}
B = U^\dagger A_1 W \Sigma^{-1},
\label{eq:B_def_clean2}
\end{equation}

\noindent and using \eqref{eq:A0A1_fact_clean}, we obtain

\begin{equation}
\begin{aligned}
B
&= U^\dagger (\Phi \Lambda C)\, W \Sigma^{-1} \\
&= (U^\dagger \Phi)\, \Lambda\, (C W \Sigma^{-1}) \\
&= S\, \Lambda\, (C W \Sigma^{-1}).
\end{aligned}
\end{equation}

\noindent Multiplying $A_0 = \Phi C$ on the left by $U^\dagger$ and using \eqref{eq:A0A1_fact_clean}, \eqref{eq:A0_svd_thin_clean} we get

\begin{equation}
U^\dagger A_0 = U^\dagger \Phi C
\quad\Longrightarrow\quad
\Sigma W^\dagger = S C.
\end{equation}

\noindent Right multiplication by $W \Sigma^{-1}$ yields

\begin{equation}
I_r = S\, C\, W\, \Sigma^{-1},
\qquad\text{hence}\qquad
C W \Sigma^{-1} = S^{-1}.
\end{equation}

\noindent Therefore,

\begin{equation}
B = S\, \Lambda\, S^{-1},
\label{eq:B_similar_clean2}
\end{equation}

\noindent showing that the diagonal matrix $\Lambda$ we are seeking is available by simply diagonalizing the reduced matrix $B$, whose construction is relatively straightforward. To accelerate the evaluation of $T(z)$ along the contour, matrix elements are
interpolated using Chebyshev expansions constructed along radial ``rays'' in
the complex plane (Fig.~\ref{fig:beyn_contour_sketch}) \cite{hank106}. Acceptance
intervals are introduced to maintain uniform interpolation accuracy in the
presence of highly oscillatory Hankel kernels. In particular, for large
arguments the outgoing Hankel function admits the asymptotic form \cite{abramowitz}

\begin{equation}
\begin{aligned}
H_{1}^{(1)}(kr)\;&\sim\;\sqrt{\frac{2}{\pi\,kr}}\,
\exp\!\left[i\left(kr-\frac{3\pi}{4}\right)\right] \\
&=\;\sqrt{\frac{2}{\pi\,kr}}\,
\exp\!\left[i\left(\Re(k)\,r-\frac{3\pi}{4}\right)\right]\,
\exp\!\big(-\Im(k)\,r\big),
\end{aligned}
\label{eq:H1_asymp}
\end{equation}

\noindent valid for $|kr|\to\infty$ where the complex wavenumber is written as $k=\Re(k)+i\,\Im(k)$. Equation~\eqref{eq:H1_asymp} shows that for complex wavenumbers the kernel acquires an exponential factor $\exp(-\Im(k)\,r)$. For $\Im(k)<0$, this term grows as $\exp(|\Im(k)|\,r)$, imposing practical constraints on the contour shape (for instance, one may employ an elliptic contour whose minor axis is oriented perpendicular to the real axis and chosen significantly smaller than the major axis, the latter determining the expected number of enclosed eigenvalues) and on the maximum distance scale within the billiard for a prescribed numerical accuracy.

\section{Symmetry reduced BIE formalism}

\subsection{Desymmetrized Fredholm matrix}
\label{irrep_bie_derivation}

Here we derive the symmetric reduced BIE. We start by defining the action of an operator $\mathbf{R}$ (representing e.g. rotation, reflection) on an arbitrary function of a vector variable $f(\mathbf{x})$ as the pullback $O_{\mathbf{R}}$

\begin{equation} \label{eq:pullback}
    O_\mathbf{R} (f(\mathbf{x})) = f(\mathbf{R^{-1}x}).
\end{equation}

\noindent For notational convenience we define the normal derivative of the Green function as $K(\mathbf{x},\mathbf{y})$ (see \eqref{eq:DLP_kernel})

\begin{equation} \label{eq:DLP_kernel_2}
    \partial_{n_{\mathbf{y}}} G_{k}(\mathbf{x},\mathbf{y}) 
    = K(\mathbf{x},\mathbf{y}) = - \frac{ik}{4}\frac{\mathbf{n(y)}\cdot(\mathbf{x}-\mathbf{y})}{d(\mathbf{x},\mathbf{y})}H_{1}^{(1)}(kd(\mathbf{x},\mathbf{y})).
\end{equation} 

\noindent We would now like to show that it is left invariant under the action of $\mathbf{R}$, that is 

\begin{equation} \label{DLP_invariance_wts}
    O_{\mathbf{R}}(K(\mathbf{x},\mathbf{y})) = K(\mathbf{R^{-1}}\mathbf{x},\mathbf{R^{-1}}\mathbf{y}) = K(\mathbf{x},\mathbf{y}),
\end{equation}

\noindent which we now prove. Writing the action of $\mathbf{R}$ explicitely

\begin{equation} \label{DLP_invariance}
\begin{aligned}
    K(\mathbf{R^{-1}}\mathbf{x},\mathbf{R^{-1}}\mathbf{y}) = - \frac{ik}{4}\frac{\mathbf{n(R^{-1}y)}\cdot(\mathbf{R^{-1}x}-\mathbf{R{-1}y})}{d(\mathbf{R^{-1}x},\mathbf{R^{-1}y})} \times \\H_{1}^{(1)}(kd(\mathbf{R^{-1}x},\mathbf{R^{-1}y})),
\end{aligned}
\end{equation}

\noindent we can use the fact that the dot product between vectors is invariant to a simultaneous rotation of both

\begin{equation} \label{rotation_dot_product}
    \mathbf{x} \cdot \mathbf{y} = (\mathbf{R^{-1}x}) \cdot (\mathbf{R^{-1}y}).
\end{equation}

\noindent Both the numerator and the denominator prefactors to the Hankel function are dot products of vectors so they are left invariant

\begin{equation}
\label{eucl_dist_invariance}
\begin{aligned}
d(\mathbf{R}^{-1}\mathbf{x},\mathbf{R}^{-1}\mathbf{y})
&=\bigl|\mathbf{R}^{-1}\mathbf{x}-\mathbf{R}^{-1}\mathbf{y}\bigr|\\
&=\bigl|\mathbf{R}^{-1}(\mathbf{x}-\mathbf{y})\bigr|\\
&=\sqrt{\bigl(\mathbf{R}^{-1}(\mathbf{x}-\mathbf{y})\bigr)\cdot
\bigl(\mathbf{R}^{-1}(\mathbf{x}-\mathbf{y})\bigr)}\\
&=\sqrt{(\mathbf{x}-\mathbf{y})\cdot(\mathbf{x}-\mathbf{y})}
\\
&=|\mathbf{x}-\mathbf{y}|\\
&=d(\mathbf{x},\mathbf{y}).
\end{aligned}
\end{equation}

\noindent The same follows for the numerator

\begin{equation}
\label{numerator_invariance}
\begin{aligned}
\mathbf{n}(\mathbf{R}^{-1}\mathbf{y})\cdot\bigl(\mathbf{R}^{-1}\mathbf{x}-\mathbf{R}^{-1}\mathbf{y}\bigr)
&=\mathbf{n}(\mathbf{R}^{-1}\mathbf{y})\cdot\mathbf{R}^{-1}(\mathbf{x}-\mathbf{y})\\
&=\bigl(\mathbf{R}^{-1}\mathbf{n}(\mathbf{y})\bigr)\cdot\mathbf{R}^{-1}(\mathbf{x}-\mathbf{y})\\
&=\bigl(\mathbf{R}^{-1}\mathbf{n}(\mathbf{y})\bigr)\cdot\bigl(\mathbf{R}^{-1}(\mathbf{x}-\mathbf{y})\bigr)\\
&=\mathbf{n}(\mathbf{y})\cdot(\mathbf{x}-\mathbf{y}),
\end{aligned}
\end{equation}

\noindent where in line $2$ we have used the fact that normals transform as

\begin{equation} \label{normal_transformation}
     \mathbf{n}(\mathbf{R}^{-1} \mathbf{y}) = \mathbf{R^{-1}n(y)}.
\end{equation}

\noindent Now using \eqref{normal_transformation} and \eqref{eucl_dist_invariance} in \eqref{DLP_invariance} it automatically follows that \eqref{eq:DLP_kernel_2} is invariant under $\mathbf{R}$ and \eqref{DLP_invariance_wts} is proven. 

\noindent Using these prerequisites the following relation 

\begin{equation} \label{eq:bie_as_A}
    A(u(\mathbf{x})) = \frac{1}{2}u(\mathbf{x}) + \int_{\Gamma_{full}} K(\mathbf{x},\mathbf{y})u(\mathbf{y})ds(\mathbf{y}) = 0,
\end{equation}

\noindent commutes with the action of $\mathbf{R}$, that is

\begin{equation} \label{eq:A_R_commutation_wts}
    [A,O_{\mathbf{R}}]u(\mathbf{x}) = A (O_{\mathbf{R}} u(\mathbf{x})) -O_{\mathbf{R}} A(u(\mathbf{x})) = 0.
\end{equation}

\noindent To show this we will need the following relations, which follow from \eqref{DLP_invariance_wts} and basic geometry

\begin{equation}
\label{eq:ds_K_symmetry}
\begin{aligned}
K(\mathbf{x},\mathbf{R}\mathbf{y})
&=K(\mathbf{R}^{-1}\mathbf{x},\mathbf{R}^{-1}\mathbf{R}\mathbf{y})
=K(\mathbf{R}^{-1}\mathbf{x},\mathbf{y}),\\
ds(\mathbf{R^{-1}}\mathbf{y})
&=ds(\mathbf{y}).
\end{aligned}
\end{equation}

Up to this point the symmetry operation $\mathbf{R}$ was arbitrary, but now we will restict ourselves only to true symmetries of the boundary $\Gamma_{full}$, since we will require $O_{\mathbf{R}}\Gamma_{full} = \Gamma_{full}$. Explicitely for a rotation this statement says

\begin{equation}
\begin{aligned}
O_{\mathbf{R}}\Gamma_{\mathrm{full}}
&=\{(r(\phi+\theta_R),\phi+\theta_R)\}\\
&=\{(r(\phi),\phi)\}
=\Gamma_{\mathrm{full}},
\end{aligned}
\end{equation}

\noindent where $\theta_R=\mathrm{angle}(R)$. Writing the action of $\mathbf{R}$ on $u(\mathbf{x})$ via \eqref{eq:pullback} and using \eqref{DLP_invariance_wts} and \eqref{eq:ds_K_symmetry} we can write

\begin{align}
A(O_{\mathbf R}u(\mathbf x))
&= \frac12\,u(\mathbf R^{-1}\mathbf x)
+ \int_{\Gamma_{\mathrm{full}}}
K(\mathbf x,\mathbf y)\,u(\mathbf R^{-1}\mathbf y)\,ds(\mathbf y)
\notag\\
&\overset{\mathbf z=\mathbf R^{-1}\mathbf y}{=}
\frac12\,u(\mathbf R^{-1}\mathbf x)
+ \int_{\Gamma_{\mathrm{full}}}
K(\mathbf x,\mathbf R\mathbf z)\,u(\mathbf z)\,ds(\mathbf z)
\notag\\
&=
\frac12\,u(\mathbf R^{-1}\mathbf x)
+ \int_{\Gamma_{\mathrm{full}}}
K(\mathbf R^{-1}\mathbf x,\mathbf z)\,u(\mathbf z)\,ds(\mathbf z) \notag\\
&= (O_{\mathbf R}A u)(\mathbf x),
\label{eq:bie_as_A_commutation_proof}
\end{align}

\noindent proving that \eqref{eq:A_R_commutation_wts} and implying that if $u(\mathbf{x})$ is a solution of \eqref{eq:bie_as_A}, then $O_{\mathbf{R}}u(\mathbf{x})=u(\mathbf{R^{-1}x})$ is also a solution. This property has an important consequence. Since the solution space is closed
under the action of the symmetry operator $O_{\mathbf R}$, solutions may be
chosen to possess definite transformation behaviour under symmetry operations.

For the rotational symmetry group $C_3=\{I,\mathbf R,\mathbf R^2\}$ one has

\begin{equation}
O_{\mathbf R}^3 u(\mathbf{x}) = u(\mathbf{x}),
\end{equation}

\noindent which follows directly from the cyclic nature of the symmetry and $\mathbf{R}$ is the rotation operator in $\mathbb{R}^2$

\begin{equation}
\mathbf{R}\,\mathbf{x} =
\begin{pmatrix}
\cos(2\pi/3) & -\sin(2\pi/3)\\
\sin(2\pi/3) & \cos(2\pi/3)
\end{pmatrix}\mathbf{x}.
\label{eq:rotation_operator}
\end{equation}

\noindent The operator $O_{\mathbf R}$ therefore admits eigenfunctions. Denoting such
eigenfunctions by $u_m(\mathbf{x})$, we define

\begin{equation}
\label{eq:u_eigen}
O_{\mathbf R} (u_m(\mathbf{x}))
=
u_m(\mathbf R^{-1}\mathbf{x})
=
\lambda_m\, u_m(\mathbf{x}).
\end{equation}

\noindent Applying the symmetry operation three times yields

\begin{equation}
(O_{\mathbf R}^3 u_m)(\mathbf{x})
=
\lambda_m^3\, u_m(\mathbf{x}).
\end{equation}

\noindent Consistency with the group property $O_{\mathbf R}^3 = I$
therefore requires

\begin{equation}
\lambda_m^3 = 1.
\end{equation}

\noindent The eigenvalues are thus the cubic roots of unity,

\begin{equation}
\lambda_m = e^{-2\pi i m/3}, \qquad m=0,1,2,
\end{equation}

\noindent and they can be identified exactly with characters of the irreducible representations,

\begin{equation} \label{eq:character_exp}
\chi_m(\mathbf R^\ell) = e^{-2\pi i m\ell/3}.
\end{equation}

\noindent Equation~\eqref{eq:u_eigen} may therefore be written as the symmetry transformation law $u_m(\mathbf R^{-1}\mathbf{x}) = \chi_m(\mathbf R)\, u_m(\mathbf{x})$, or more generally,

\begin{equation}
\label{eq:u_transform_general}
u_m(\mathbf R^{-\ell}\mathbf{x})
=
\chi_m(\mathbf R^\ell)\, u_m(\mathbf{x}).
\end{equation}

\noindent Physically, this expresses that boundary functions can be classified according to their symmetry behaviour. Each solution belongs to a definite symmetry sector. To extract solutions with definite symmetry, we introduce the projection
operator onto the sector $m$,

\begin{equation}
\label{eq:projector_u}
u_m(\mathbf{x}) = P_m u(\mathbf{x}) = \frac{1}{3} \sum_{\ell=0}^{2} \chi_m(\mathbf R^\ell)^{*}\, O_{\mathbf R}^\ell u(\mathbf{x}).
\end{equation}

\noindent Inserting $u_m(\mathbf{x})$ into \eqref{eq:bie_as_A} we can see that it is indeed a solution whenever $u(\mathbf{x})$ is

\begin{equation}
A (u_m(\mathbf{x})) = A (P_m u(\mathbf{x})) = P_m ( A (u(\mathbf{x})) = 0,
\end{equation}

\noindent where we have used the fact that $P_m$ is a linear operator. Consequently, the boundary integral equation decomposes into independent symmetry sectors
labelled by $m$. This allows the desymmetrization of \eqref{eq:bie_as_A} into an integral over only the fundamental domains if we restrict ourselves to only solutions $u_m(\mathbf{x})$ in a given irreducible representation labeled with $m=0,1,2$. Therefore we decompose the integration over the full boundary into an integral over the fundamental domain and the other parts of the domain will be constructed from symmetry

\begin{flalign}
&\int_{\Gamma_{\mathrm{full}}} K(\mathbf{x},\mathbf{y})\,u_m(\mathbf{y})\,ds(\mathbf{y})=\notag &\\
&=\sum_{\ell=0}^{2}\int_{\Gamma_{\mathrm{fund}}}
K(\mathbf{x},\mathbf{R}^{\ell}\mathbf{y})\,u_m(\mathbf{R}^{\ell}\mathbf{y})\,ds(\mathbf{y})\notag &\\
&=\sum_{\ell=0}^{2}\chi_m(\mathbf{R}^{-\ell})
\int_{\Gamma_{\mathrm{fund}}}
K(\mathbf{x},\mathbf{R}^{\ell}\mathbf{y})\,u_m(\mathbf{y})\,ds(\mathbf{y})\notag &\\
&=\sum_{\ell=0}^{2}\chi_m(\mathbf{R}^{\ell})^{*}
\int_{\Gamma_{\mathrm{fund}}}
K(\mathbf{x},\mathbf{R}^{\ell}\mathbf{y})\,u_m(\mathbf{y})\,ds(\mathbf{y}).
\label{desymmetrized_integral} &
\end{flalign}

\noindent Finally inserting \eqref{desymmetrized_integral} into \eqref{eq:bie_as_A} we have the equation 

\begin{equation} \label{eq:bie_as_A_desymmetrized}
\begin{aligned}
    A(u_m(\mathbf{x})) = \frac{1}{2}u_m(\mathbf{x}) + \\ \sum_{\ell=0}^{2}\chi_m(\mathbf{R}^{\ell})^{*} \int_{\Gamma_{\mathrm{fund}}} K(\mathbf{x},\mathbf{R}^{\ell}\mathbf{y})\,u_m(\mathbf{y})\,ds(\mathbf{y}).
\end{aligned}
\end{equation}

\noindent Upon discretization of the fundamental boundary 
$\Gamma_{\mathrm{fund}}$ using quadrature nodes 
$\{\mathbf{x}_i\}_{i=1}^{N}$ and quadrature weights $\{\zeta_j\}_{j=1}^{N}$ (e.g. Gauss-Legendre), the BIE reduces to a matrix equation. The
symmetry--reduced boundary integral equation becomes

\begin{equation}
\label{eq:matrix_equation}
\sum_{j=1}^{N}
A^{(m)}_{ij}\,u_m(\mathbf{x}_j)=0,
\end{equation}

\noindent where the matrix elements are given by

\begin{equation}
\label{eq:matrix_elements}
A^{(m)}_{ij} =
\frac{1}{2}\delta_{ij} + \sum_{\ell=0}^{2} \chi_m(\mathbf{R}^{\ell})^{*}\, \zeta_j\, K(\mathbf{x}_i,\mathbf{R}^{\ell}\mathbf{x}_j),
\end{equation}

\noindent The matrix \eqref{eq:matrix_elements} is then used in Beyn's method ($T(z)$ matrix as used in \eqref{eq:keldysh}--\eqref{eq:contour_moments}) as described in Appendix.\ref{app:beyn}.

\subsection{Wavefunction in a given irreducible representation}
\label{irrep_wavefunction_derivation}

For Dirichlet boundary conditions $\psi|_{\Gamma_{\mathrm{full}}}=0$, Green's
theorem reduces to a pure single-layer potential term. Hence the interior
wavefunction in symmetry sector $m$ is reconstructed as ($G_k(\mathbf{x},\mathbf{y})$ given in \eqref{eq:green_function})
\begin{equation}
\label{eq:psi_slp_full}
\psi_m(\mathbf{x})
= -
\int_{\Gamma_{\mathrm{full}}}
G_k(\mathbf{x},\mathbf{y})\,u_m(\mathbf{y})\,ds(\mathbf{y}),
\qquad \mathbf{x}\in\Omega .
\end{equation}

\noindent Following the procedure above we reduce the boundary integral to the fundamental boundary segment
$\Gamma_{\mathrm{fund}}$ using $\Gamma_{\mathrm{full}}=\bigcup_{\ell=0}^{2}\mathbf{R}^{\ell}\Gamma_{\mathrm{fund}}$
and \eqref{eq:ds_K_symmetry}, obtaining
\begin{equation}
\label{eq:psi_slp_split}
\psi_m(\mathbf{x})
=
-
\sum_{\ell=0}^{2}
\int_{\Gamma_{\mathrm{fund}}}
G_k(\mathbf{x},\mathbf{R}^{\ell}\mathbf{y})\,
u_m(\mathbf{R}^{\ell}\mathbf{y})\,ds(\mathbf{y}).
\end{equation}

\noindent Using the symmetry transformation law \eqref{eq:u_transform_general}, i.e.
$u_m(\mathbf{R}^{\ell}\mathbf{y})=\chi_m(\mathbf{R}^{-\ell})u_m(\mathbf{y})
=\chi_m(\mathbf{R}^{\ell})^{*}u_m(\mathbf{y})$, we arrive at the symmetry-reduced
reconstruction formula

\begin{equation}
\label{eq:psi_slp_fund}
\psi_m(\mathbf{x})
=
\textcolor{red}{-}
\sum_{\ell=0}^{2}\chi_m(\mathbf{R}^{\ell})^{*}
\int_{\Gamma_{\mathrm{fund}}}
G_k(\mathbf{x},\mathbf{R}^{\ell}\mathbf{y})\,
u_m(\mathbf{y})\,ds(\mathbf{y}),
\end{equation}

\noindent where numerically the boundary integral over $\Gamma_{fund}$ is discretized using the same procedure as going from \eqref{eq:bie_as_A_desymmetrized} to \eqref{eq:matrix_elements}.

\subsection{Poincar\'e--Husimi (PH) function in a given irreducible representation}
\label{irrep_ph_derivation}

The Poincar\'e--Husimi function (unnormalized) associated with a boundary density is defined as \cite{husimi_original,Backer2003}

\begin{equation}
H(q,p)=\bigl|\langle c_{q,p}\mid u\rangle\bigr|^2,
\qquad
\langle c_{q,p}\mid u\rangle
=
\int_{\Gamma_{\mathrm{full}}}
c_{q,p}^*(\mathbf{y})\,u(\mathbf{y})\,ds(\mathbf{y}),
\label{eq:backer_boundary_husimi}
\end{equation}
\noindent where $c_{q,p}(\mathbf{y})$ denotes a boundary coherent state localized at
the phase--space point $(q,p)$. Restricting to a symmetry sector and using the boundary decomposition $\Gamma_{\mathrm{full}}=\bigcup_{\ell=0}^{2}\mathbf R^{\ell}\Gamma_{\mathrm{fund}}$
together with the invariance $ds(\mathbf R^{\ell}\mathbf y)=ds(\mathbf y)$,
the projection amplitude becomes

\begin{equation}
\begin{aligned}
\langle c_{q,p}\mid u_m\rangle
&=
\sum_{\ell=0}^{2}
\int_{\Gamma_{\mathrm{fund}}}
c_{q,p}^*(\mathbf R^{\ell}\mathbf y)\,
u_m(\mathbf R^{\ell}\mathbf y)\,ds(\mathbf y) \\
&=
\sum_{\ell=0}^{2}
\chi_m(\mathbf R^{\ell})^{*}
\int_{\Gamma_{\mathrm{fund}}}
c_{q,p}^*(\mathbf R^{\ell}\mathbf y)\,
u_m(\mathbf y)\,ds(\mathbf y),
\end{aligned}
\label{eq:ph_symmetry_reduced_amp}
\end{equation}
where Eq.~\eqref{eq:u_transform_general} has been used. The PH function in the $m$-th irreducible representation therefore reads

\begin{equation}
H_m(q,p)
=
\left|
\sum_{\ell=0}^{2}
\chi_m(\mathbf R^{\ell})^{*}
\int_{\Gamma_{\mathrm{fund}}}
c_{q,p}^*(\mathbf R^{\ell}\mathbf y)\,
u_m(\mathbf y)\,ds(\mathbf y)
\right|^2 .
\label{eq:ph_final}
\end{equation}

\end{document}